\documentclass[9pt,twoside]{rmaa-rho-class/rmaa-rho}
\RMxAAtemplatetype{\RMxAA} % Select your article type

\vol{100}
\pages{1000-1025}
\thisyear{2026}
\doi{\href{https://doi.org/10.22201/ia.01851101p.20XX.XX.XX.XX}{https://doi.org/10.22201/ia.01851101p.20XX.XX.XX.XX}}

\usepackage{tikz}
\usetikzlibrary{shapes.geometric, arrows.meta, positioning}
\usepackage{amsmath}
\usepackage{ulem}

\def\ha{{\sc H}$\alpha$}
\def\oiii{{\sc [O~III]}$\lambda$5007}

\def\sii{{\sc [S~II]}$\lambda\lambda$6716,6731}

\usepackage{graphicx}	% Including figure files
\usepackage{amsmath}	% Advanced maths commands

\title{SDSS-V Local Volume Mapper (LVM): Dithered Data Cube Reconstruction with {\sc 3dcubegen}}
\author[1]{H.~Ibarra-Medel \orcidlink{0000-0002-9790-6313}}
\author[2,3]{A. Z. Lugo-Aranda\orcidlink{0000-0001-9226-9178}}
\author[1]{R. de J. Zermeño\orcidlink{0009-0009-0081-4323}}
\author[1,4]{Sebastián F. Sánchez\orcidlink{0000-0001-6444-9307}}
\author[1]{Lesly Castañeda-Carlos\orcidlink{0009-0000-2341-9865}}
\author[1]{J. Eduardo Méndez-Delgado\orcidlink{0000-0002-6972-6411}}
\author[6]{K. Kreckel\orcidlink{0000-0001-6551-3091}}
\author[7]{Roeland P. van der Marel\orcidlink{0000-0001-7827-7825}}
\author[11]{R. Orozco-Duarte \orcidlink{0000-0003-2193-3005}}
\author[2]{Wofford, Aida\orcidlink{0000-0001-8289-3428}}
\author[1,5]{Castalia Alenka Negrete\orcidlink{0000-0002-1656-827X}}
\author[1]{Irene Cruz-Gonz\'{a}lez\orcidlink{0000-0002-2653-1120}}
\author[2]{Carlos G. Román-Zúñiga \orcidlink{0000-0001-8600-4798}}
\author[12,13]{Guillermo A. Blanc \orcidlink{0000-0003-4218-3944}}
\author[9]{Evelyn J. Johnston \orcidlink{0000-0002-2368-6469}}
\author[8]{Ivan Katkov \orcidlink{0000-0002-6425-6879}}
\author[10]{Alfredo J Mejía-Narváez \orcidlink{0000-0002-8931-2398}}
\author[14]{Tony Wong \orcidlink{0000-0002-0786-7307}}
\author[15]{Oleg~V.~Egorov \orcidlink{0000-0002-4755-118X}}

\affil[1]{Universidad Nacional Autónoma de México. Instituto de Astronomía. A.P. 70-264, 04510. Ciudad de México, México.}
\affil[2]{Universidad Nacional Autónoma de México. Instituto de Astronomía. A.P. 106, 22800. Ensenada, B.C. , México}
\affil[3]{Institute of Astrophysics, Facultad de Ciencias Exactas, Universidad Andrés Bello, Sede Concepción, Talcahuano, Chile}
\affil[4]{Instituto de Astrofísica de Canarias, Vía Láctea s/n, 38205, La Laguna, Tenerife, Spain}
\affil[5]{Cátedras CONAHCYT. Universidad Nacional Autónoma de México, Instituto de Astronomía, A.P. 70-264, 04510, Ciudad de México, México}
\affil[6]{Astronomisches Rechen-Institut, Zentrum f\"{u}r Astronomie der Universit\"{a}t Heidelberg, M\"{o}nchhofstra\ss e 12-14, D-69120 Heidelberg, Germany}
\affil[7]{Space Telescope Science Institute, 3700 San Martin Drive, Baltimore, MD 21218, USA}
\affil[8]{New York University Abu Dhabi, PO Box 129188, Abu Dhabi, UAE}
\affil[9]{Universidad Diego Portales, Instituto de Estudios Astrofísicos, Facultad de Ingeniería y Ciencias, Av. Ejército Libertador 441, Santiago, Chile}
\affil[10]{Universidad de Chile, Av. Libertador Bernardo O'Higgins 1058, Santiago de Chile}
\affil[11]{Instituto de Radioastronomía y Astrofísica, Universidad Nacional Autónoma de México, 58090 Morelia, Michoacán, México}
\affil[12]{Observatories of the Carnegie Institution for Science, 813 Santa Barbara Street, Pasadena, CA 91101, USA}
\affil[13]{Departamento de Astronom\'{i}a, Universidad de Chile, Camino del Observatorio 1515, Las Condes, Santiago, Chile}
\affil[14]{Department of Astronomy, University of Illinois at Urbana-Champaign, Urbana, IL 61801, USA}
\affil[15]{Astronomisches Rechen-Institut, Zentrum f\"ur Astronomie der Universit\"at Heidelberg, M\"onchhofstr. 12-14, D-69120 Heidelberg, Germany.}

\leadauthor{Ibarra-Medel et al.}
\smalltitle{Data Cube Reconstruction}

\corres{Hector Ibarra-Medel (HIM)}
\email{hibarram@astro.unam.mx}

\received{April 16th, 2024}
\accepted{\today}
\license{Texto de la licencia aquí}

\setbool{rho-abstract}{true} % Set false to hide the abstract
\setbool{rho-resumen}{true} % Set false to hide the abstract

\begin{abstract}
The Sloan Digital Sky Survey V (SDSS-V) Local Volume Mapper (LVM) is conducting an unprecedented wide-field integral field spectroscopic survey of the Milky Way, the Magellanic Clouds, and nearby galaxies using a strategy based on multiple dithered observations to achieve full spatial coverage, improved spatial sampling, and enhanced spectral depth. However, the scientific exploitation of these observations requires a robust methodology to combine the individual row-stacked spectra (RSS) into homogeneous three-dimensional data cubes. In this work, we present 3DCubeGen, a flexible and scalable reconstruction tool designed to combine multiple LVM dithers while preserving flux and propagating uncertainties. The method enables the coaddition of large datasets, improving the signal-to-noise ratio, enhancing spatial resolution, and increasing sensitivity to faint emission features, following and extending approaches previously implemented in integral field surveys such as CALIFA. We apply 3DCubeGen to a large set of LVM observations, including the Large and Small Magellanic Clouds and nearby galaxies, combining thousands of dithers corresponding to millions of spectra. The resulting data products demonstrate significant improvements in spatial sampling and spectral depth, enabling detailed studies of the ionised gas, stellar populations, and kinematics across extended regions. 3DCubeGen provides a robust and scalable solution for LVM data cube reconstruction and represents a key tool for exploiting the scientific potential of the SDSS-V Local Volume Mapper.
\end{abstract}


\begin{resumen}
El Local Volume Mapper (LVM) del Sloan Digital Sky Survey V (SDSS-V) está llevando a cabo un estudio espectroscópico de campo integral sin precedentes del plano de la Vía Láctea, las Nubes de Magallanes y galaxias cercanas, utilizando una estrategia basada en múltiples observaciones con dithering para lograr cobertura espacial completa, mejorar el muestreo espacial y aumentar la profundidad espectral. Sin embargo, la explotación científica de estas observaciones requiere una metodología robusta para combinar los espectros apilados por filas (RSS) individuales en cubos de datos tridimensionales homogéneos. En este trabajo presentamos 3DCubeGen, una herramienta de reconstrucción flexible y escalable diseñada para combinar múltiples dithers del LVM preservando el flujo y propagando las incertidumbres. Este método permite la coadición de grandes volúmenes de datos, mejorando la relación señal-ruido, aumentando la resolución espacial y la sensibilidad a características de emisión débiles, siguiendo y extendiendo enfoques previamente implementados en estudios de espectroscopía de campo integral como CALIFA. Aplicamos 3DCubeGen a un amplio conjunto de observaciones del LVM, incluyendo la Gran y la Pequeña Nube de Magallanes y galaxias cercanas, combinando miles de dithers que corresponden a millones de espectros. Los cubos de datos resultantes muestran mejoras significativas en el muestreo espacial y la profundidad espectral, permitiendo estudios detallados del gas ionizado, las poblaciones estelares y la cinemática en regiones extensas. 3DCubeGen proporciona así una solución robusta y escalable para la reconstrucción de cubos de datos del LVM y representa una herramienta clave para explotar el potencial científico del Local Volume Mapper del SDSS-V.
\end{resumen}

\begin{document}

\maketitle
\pagestyle{fancy}\thispagestyle{firststyle}%

\section{Introduction}

Over the past two decades, Integral Field Spectroscopy (IFS) has transformed the way we study the physical properties of galaxies in both galactic and extragalactic contexts \citet{Bacon+2010,Sanchez+2020}. This technique enables spatially resolved analysis of spectral features, offering detailed insights into the physical properties of ionised gas, chemical enrichment, stellar kinematics, and stellar populations \citep[see ][and references therein]{Cappellari+2011,Perez+2013,Gonzalez-Delgado+2017,Aquino+2020,Cano-Diaz+22,Artemi+2022,Avila-Reese+2023,Kreckel+2024,Ibarra-Medel+2016,Ibarra-Medel+2022a,Ibarra-Medel+2025}. This progress has led to major IFS galaxy surveys such as the Calar Alto Legacy Integral Field Area survey \citep[CALIFA,][]{Sanchez+2012}, the Sydney-AAO Multi-object Integral field spectrograph survey \citep[SAMI,][]{Croom+2012}, and the Mapping Nearby Galaxies at Apache Point Observatory survey \citep[MaNGA]{Bundy+2015}, part of the fourth phase of the Sloan Digital Sky Survey \citep[SDSS-IV,][]{Abdurro+22}. These galaxy surveys explored the nearby universe ($0.01 \leq z \leq 0.05$), enabling spatially resolved mapping of galaxy properties at scales of $\simeq$ 1 kpc. In addition, advances in new IFS instrumentation such as MUSE \citep{Bacon+2010} and SITELLE \citep{Grandmont+2012} have pushed the limits of IFS to achieve seeing-limited spectroscopy.

%Looking ahead, the Local Volume Mapper (LVM) from the fifth phase of the Sloan Digital Sky Survey (SDSS-V) further expands the capabilities of IFS observations. SDSS-V is the first all-sky multi-epoch and integral field spectroscopic survey \citep[][]{Kollmeier+2026,Almeida+2023}. SDSS-V comprises three core projects or mappers: the Black Hole Mapper (BHM), the Milky Way Mapper (MWM), and the Local Volume Mapper (LVM). The BHM and the MWM use optical/IR multi-epoch spectroscopy and map the two hemispheres simultaneously with the 2.5-m telescope at Apache Point Observatory (APO) in the northern hemisphere and with the 100-inch du-Pont Telescope at Las Campanas Observatory (LCO) in the southern hemisphere. The third mapper, the LVM, is creating the most ambitious integral field spectroscopic survey of the Milky Way (MW), the Magellanic Clouds (MC), and Local Volume (LV) galaxies. The LVM has designed and constructed a new robotic observatory at LCO for this purpose \citep{Drory+2024}. The LVM will map the southern part of the MW at spatial resolutions from 0.05 to 1 pc, the MCs at $\sim$4 pc resolution, and 423 LV galaxies, covering more than 43,000 deg$^2$ with IFS observations \citep{Drory+2024}.

The fifth phase of the Sloan Digital Sky Survey (SDSS-V) comprises three core projects or mappers: the Black Hole Mapper (BHM), the Milky Way Mapper (MWM), and the Local Volume Mapper (LVM) \citep[][]{Kollmeier+2026,Almeida+2023}. The BHM and the MWM use optical/IR multi-epoch spectroscopy and map the two hemispheres simultaneously with the 2.5-m telescope at Apache Point Observatory (APO) in the northern hemisphere and with the 100-inch du-Pont Telescope at Las Campanas Observatory (LCO) in the southern hemisphere. The third mapper, the LVM, is creating the most ambitious integral field spectroscopic survey of the Milky Way (MW), the Magellanic Clouds (MC), and Local Volume (LV) galaxies. The LVM has designed and constructed a new robotic observatory at LCO for this purpose \citep{Drory+2024}. The LVM will map the southern part of the MW at spatial resolutions from 0.05 to 1 pc, the MCs at $\sim$4 pc resolution, and 423 LV galaxies, covering more than 43,000 deg$^2$ with IFS observations \citep{Drory+2024}. By these means, SDSS-V further expands the capabilities of integral field spectroscopy observations.

Part of the LVM observing strategy, particularly for the MCs and LV observations, relies on multiple dithered exposures to achieve full spatial coverage and improved spatial sampling. The LVM Data Reduction Pipeline (LVM-DRP) produces fully calibrated row-stacked spectra (RSS) for each individual pointing. However, scientific exploitation of the LVM data requires combining these dithered observations to reach deeper sensitivity levels and detect faint emission lines. Therefore, a tool to combine these data is essential not only to recover spatial information but also to increase the signal-to-noise ratio through the coaddition of multiple exposures. Consequently, a robust and flexible data cube reconstruction methodology is required to fully explore the scientific potential of the LVM observations.

\section{The LVM}

The LVM robotic telescope is an innovative observing infrastructure at LCO formed by four hardware subsystems: The telescopes, the IFU and fibre systems, the spectrographs and the enclosure. For further details, see \citet{Drory+2024} and references therein \citep[see also][]{Herbst+2020,Herbst+2022,Herbst+2024,Konidaris+2020,Perruchot+2018,Feger+2020}. The LVM uses four telescopes: two dedicated to observing the sky map with a detailed sky model for background subtraction, one dedicated to observing the spectrophotometric stars for flux calibration, and only one dedicated to observing the science targets. The telescopes are 16.1 cm aperture refractive f/11.42 telescopes \citep{Lanz+2022} with a 1.4 degree Field of View. 

%The innovative design of the telescope consist on the base that the optics are mounted in an isolated optical bench \citep{Herbst+2024}, with the sky image redirected by two flat mirrors mounted into a siderostat. Therefore, this design creates fixed optics for each telescope and creates a fibre system that does not move, eliminating any flexures on the spectrographs \citep[see Figure 2 and 3 in ][]{Drory+2024}. 
At the optical plane of the science telescope, it is mounted a fibre-couple micro-lens array IFUs: the science telescope has an IFU with 1,801 fibres. Each fibre consists of 107$\mu$m core diameter optical fibre that, with the f/11.42 telescopes, gives a circular aperture of 35.3 arcsecs. Therefore, the IFU science returns a hexagon shape FoV of 15.4 arcminute sides or 30.2 arcminute diameter covering a total sky area of 0.165 $deg^2$ with an 83$\%$ of filling factor \citep{Drory+2024}. The IFU is divided into three segments for each spectrograph. Therefore, the four telescopes fed 1,944 fibres that meet in a sorting box, rearranging into three 648 fibre bundles with a mix of science, sky and spectrophotometric fibres \citep[][see also Figures 3 and 4 of \citealt{Drory+2024}]{Feger+2020}. The spectrographs are based on the Dark Energy Spectroscopic Instrument \citep[DESI,][]{Perruchot+2018}, and consist of three spectrographs with three dichroic spectral channels each, one for the blue (3,600-5,800\AA), one for the red (5,750-7,570\AA) and one for the infrared (7,520-9,800\AA). The combined channels generate an effective spectral range from 3,600 to 9,800 \AA~with a spectral resolution of $R\sim 4,000$.

\begin{figure}[t]
\centering 
\includegraphics[width=0.98\columnwidth]{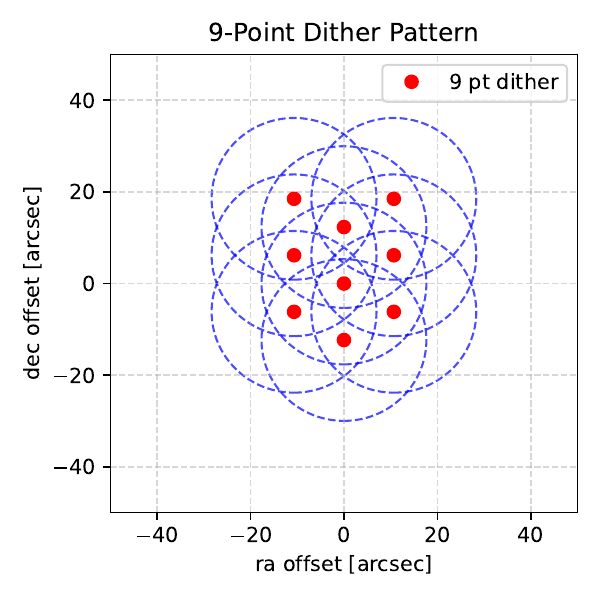}
\caption{LVM 9-dithered pattern. The red dots represents the centroids of the fibres, and the blue dashed circles represents the diameter of the fibres.}
\label{fig:dither}
\end{figure}

The selection of the tiles depended on the science schedule of the night, and depending on the object, it could be one pointing per tile (fill factor of 83$\%$) up to 9-dither pattern pointing per tile (see Figure\ref{fig:dither}), that aims to achieve a fill factor of $100\%$. Each science exposure consists of a 900 s long. The reduction of the data is carried out by the LVM Data Reduction Pipeline\footnote{\url{https://github.com/sdss/lvmdrp}} \citep[{\sc LVM-DRP}, Mejia-Narvaez in prep and][]{Sanchez+2024}. The LVM-DRP follows standard IFS reduction procedures based on the Py3D pipeline \citep{Husemann+2013}, based in turn on the methodology described in \citet{Sanchez+2006}, but it incorporates the specific requirements of the LVM data. The {\sc LVM-DRP} reduction process can be summarized as follows: 
\begin{itemize}
    \item[a)] Raw data processing to homogenize the deferents detector gains across all the blue, red, and infrared channels, followed by bias and cosmic ray correction. 
    \item[b)] Tracing each fibre spectra across each detector for all channels and spectrographs across the cross-dispersion and dispersion axes.
    \item[c)] Extraction of each spectra from the traces and applying a stray-light correction.
    \item[d)] Obtain the wavelength solution and linear resampling of each extracted spectra.
    \item[f)] Correction of the fibre-to-fibre transmission discrepancies.
    \item[g)] Combining all the channels into a single calibrated spectra.
    \item[h)] Modelling and subtracting the sky spectra using the data taken from the two sky telescopes.
    \item[i)] Implement the astrometry solution from the guide star images taken simultaneously with the observation.
\end{itemize}

Therefore, the final product of the {\sc LVM-DRP} is a row-stack-spectra \citep[RSS][]{Sanchez+2006} fits file containing all the 1,944 wavelength and flux-calibrated sky-subtracted spectra for the the IFUs of the LVM telescopes. The error propagation is performed along the reduction, generating an inverse variance RSS. In addition, the wavelength dependence on the instrumental resolution is estimated based on the line spread function (LSF) of the wavelength calibration lamps. Finally, a table with the fibre slit map is saved with the information on the position (physical and astrometric) of all science fibres with their position on the RSS file. The {\sc LVM-DRP} names the final file as the lvmCFrame per each tile/pointing.

After the {\sc LVM-DRP} produces the final reduced spectra, an additional step is carried out with the Data Analysis Pipeline\footnote{\url{https://github.com/sdss/lvmdap}} \citep[{\sc LVM-DAP},][]{Sanchez+2024}. This pipeline, which is based on {\sc pyPipe3D}\footnote{\url{https://gitlab.com/pipe3d/pyPipe3D/}} \citep{Lacerda+2022,Sanchez+2016a,Sanchez+2016b,Sanchez+2022}, performs a systematic spectral analysis that consists in the following steps: A non-linear fitting of the stellar populations, that aims to obtain an initial value of the stellar velocity shift, stellar dispersion and dust attenuation. Then, the {\sc LVM-DAP} performs an iterative process to make a parametric emission line fit to a set of strong emission lines and a stellar population decomposition using a set of resolved stellar populations to model the stellar continuum. Finally, the {\sc LVM-DAP} performs a non-parametric analysis of the stellar-free emission line spectra to retrieve the properties (velocity shifts, velocity dispersion, fluxes and equivalent widths) of the complete set of strong and weak emission lines \citep{Sanchez+2024}.

\subsection{LVM dithered observations}

The LVM observes the full LMC, SMC, and Local Volume galaxies using a 9-point dither pattern to achieve a filling factor of $100\%$ per tile. The observations of the LMC with the LVM started on 14th August 2023 (MJD 60201) during the commissioning and early science phase. The LMC was periodically observed periodically until 7th May 2024 (MJD 60437), when the LVM entered full survey mode, and later resumed from 2th July 2024 (MJD 60493) to the present. For the SMC, the first observations were obtained between 14th October 2023 (MJD 60231) and 11th November 2023 (MJD 60259) during the early science phase, and were resumed in survey mode from 2th July 2024 (MJD 60493) to the present. At the time of writing, the LVM has observed a total of 4,000 dithers for the Magellanic Clouds. However, after applying quality control criteria and considering the specific geometry required for the data cube reconstruction (see next section), and in order to maximize the number of tiles while minimizing memory requirements, we selected 705 dithers for the SMC and 1,831 dithers for the LMC. In addition, the LVM observed 26 Local Volume galaxies with a total of 806 dithers. 

Altogether, this corresponds to approximately $\sim$6,000,000 spectra used in this work. When combined, these dithered observations improve the signal-to-noise ratio, enhance spatial sampling, and increase the effective spectral depth of the data. This approach has been successfully implemented in previous integral field surveys such as CALIFA \citep{Sanchez+2012} and subsequent studies \citep[e.g.,][]{Marmol+2011}, where the combination of multiple dithered pointings enabled improved spatial resolution and increased sensitivity to faint emission features.

\section{Data cube reconstruction}\label{sec:cubrec}

For the data-cube reconstruction, we present the {\sc 3dcubegen}\footnote{\url{https://github.com/hjibarram/3DCubeGen}} python package, that is an improved version of the methodology showed in \citet{Ibarra-Medel+2019} to reconstruct the MaNGA data cubes from hydrodynamical simulations and later applied it to IFS actual data in \citet{Benitez+2023} to reconstruct the data-cubes of GTC-MEGARA observations. The {\sc 3dcubegen} methodology's core is heavily based on the method described by \citet{Sanchez+2023} for the data-cube reconstruction of the new CALIFA remastered data release. This methodology is also used for the data-cube reconstruction of the MaNGA \citep{Law+2016} and CALIFA surveys \citep{Sanchez+2012}. While the method is the same, the main difference between the MaNGA and CALIFA data cubes from the LVM dithered observations is the total number of dithers: They used a three dither pattern, while for the LVM, a new dithering was used: $(\Delta \alpha\arcsec, \Delta \delta \arcsec)=$ (0.00, 0.00), (-10.68, 18.50), (10.68, 18.50), (-0.00, -12.33), (10.68, -6.17), (-10.68, -6.17), (10.68, 6.17), (-10.68, 6.17), (0.00, 12.33) as we show in Figure~\ref{fig:dither}.

%The {\sc 3dcubegen} data-cube reconstruction consists of three principal steps: i) The 2D spatial interpolation. ii) The kernel selection, and iii) 3D cube reconstruction. We proceed to describe each step in detail:

\begin{figure}
\centering
\resizebox{\columnwidth}{!}{%
\begin{tikzpicture}[
  font=\small,
  node distance=5.5mm and 8mm,
  io/.style={trapezium, trapezium left angle=70, trapezium right angle=110,
    draw, thick, fill=gray!12, text width=52mm, align=center,
    minimum height=9mm, inner xsep=1mm},
  proc/.style={rectangle, draw, thick, fill=blue!8, text width=56mm,
    align=center, minimum height=9mm, rounded corners=1pt},
  opt/.style={rectangle, draw, thick, dashed, fill=orange!10,
    text width=34mm, align=center, minimum height=9mm, rounded corners=1pt},
  dec/.style={rectangle, draw, thick, fill=green!8, text width=56mm,
    align=center, minimum height=9mm, rounded corners=1pt},
  arr/.style={-{Stealth[length=2.4mm]}, thick},
  oarr/.style={-{Stealth[length=2.4mm]}, thick, dashed},
  lab/.style={font=\scriptsize\itshape, midway, fill=white, inner sep=1pt}
]

% --- Main vertical pipeline ---
\node[io] (input) {\textbf{Input: LVM-DRP \texttt{lvmSFrame} RSS}\\
  flux, inverse variance, slitmap\\ ($N$ dithers $\times$ tiles)};

\node[proc, below=of input] (prep) {\textbf{Pre-processing}\\
  heliocentric velocity correction;\\ aperture correction $f_A = A_{\rm spax}/A_{\rm fib}$};

\node[dec, below=of prep] (kernel) {\textbf{Kernel selection} ($\sigma$, $\alpha$; Table~\ref{tab:resolutions})\\
  truncation radius $r \leq 2\sigma$\\ (min.\ $7\,D_{\rm fib}$ for small kernels)};

\node[proc, below=of kernel] (interp) {\textbf{2D spatial interpolation} (Eq.~\ref{spatial_rs})\\
  repeated per spectral pixel\\ (3600--9800\,\AA, $\Delta\lambda = 0.5$\,\AA)};

\node[proc, below=of interp] (err) {\textbf{Error propagation} (Sec.~\ref{secc:noise})\\
  per-spaxel variance extension;\\ optional covariance tensor\\ (\texttt{fcovmat=True})};

\node[io, below=of err] (cube) {\textbf{Output: 3D data cube}\\
  flux + error FITS extensions};

% --- Optional branches ---
\node[opt, right=of interp] (maps2d) {\textbf{2D map\\ reconstruction}\\
  Eq.~\ref{spatial_rs} applied directly to\\ RSS / DAP quantities\\
  $\rightarrow$ 2D maps + error maps\\ (step iv, bypasses cube)};

\node[opt, right=of cube] (rss) {\textbf{Cube-to-RSS\\ conversion} (step v)\\
  coadded spectra +\\ astrometric fibre map};

\node[opt, below=of rss, text width=34mm] (dap) {\textbf{LVM-DAP}\\
  spectral analysis\\ (Sec.~\ref{sec:dapimp})};

% --- Arrows: main pipeline ---
\draw[arr] (input) -- (prep);
\draw[arr] (prep) -- (kernel);
\draw[arr] (kernel) -- (interp);
\draw[arr] (interp) -- (err);
\draw[arr] (err) -- (cube);

% --- Arrows: optional branches ---
\draw[oarr] (kernel.east) -| (maps2d.north) node[lab, pos=0.75, right=1pt]{optional};
\draw[oarr] (cube.east) -- (rss.west) node[lab]{optional};
\draw[oarr] (rss) -- (dap);

\end{tikzpicture}%
}
\caption{Workflow of the \textsc{3dcubegen} data cube reconstruction. Solid boxes trace the main pipeline from the LVM-DRP row-stacked spectra to the final data cube: pre-processing of the input spectra, kernel selection, 2D spatial interpolation repeated per spectral pixel, and error propagation. Dashed boxes indicate the two optional operating modes: the direct 2D map reconstruction of RSS- or DAP-derived quantities (step iv of Section~\ref{sec:cubrec}), which bypasses the full cube reconstruction, and the cube-to-RSS conversion (step v) that produces spatially co-added spectra for ingestion into the LVM-DAP.}
\label{fig:workflow}
\end{figure}
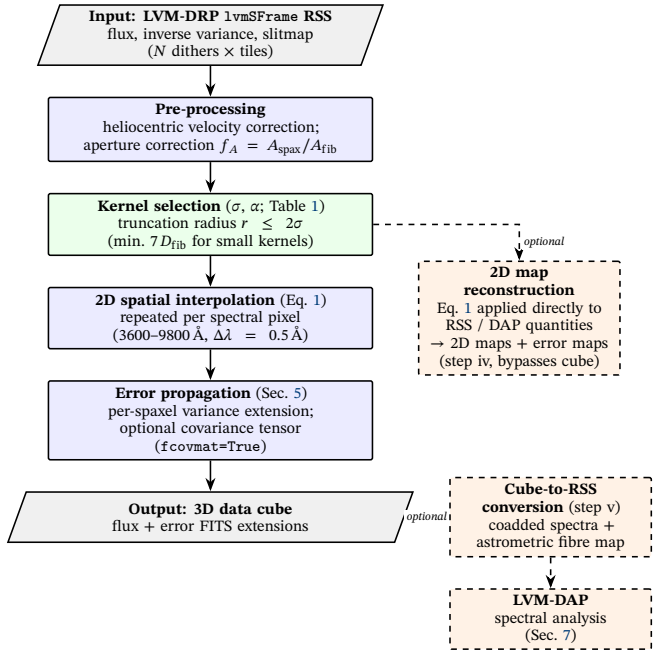

\begin{figure*}[!t]
\centering 
\includegraphics[width=2\columnwidth]{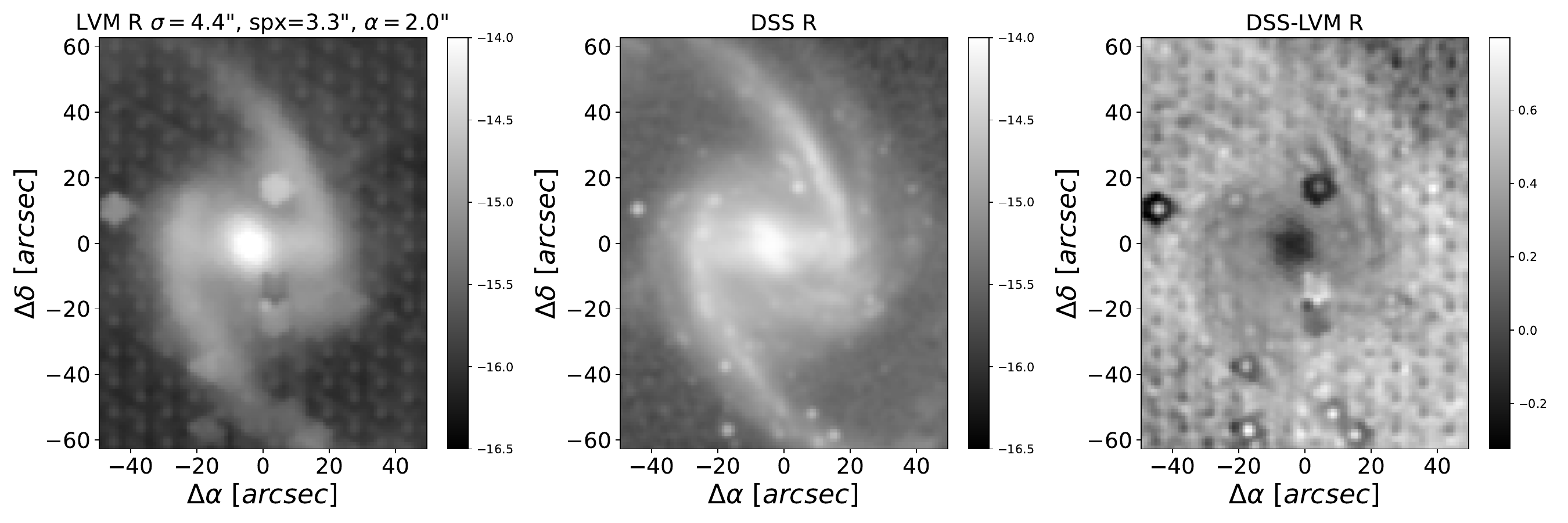}
\includegraphics[width=2\columnwidth]{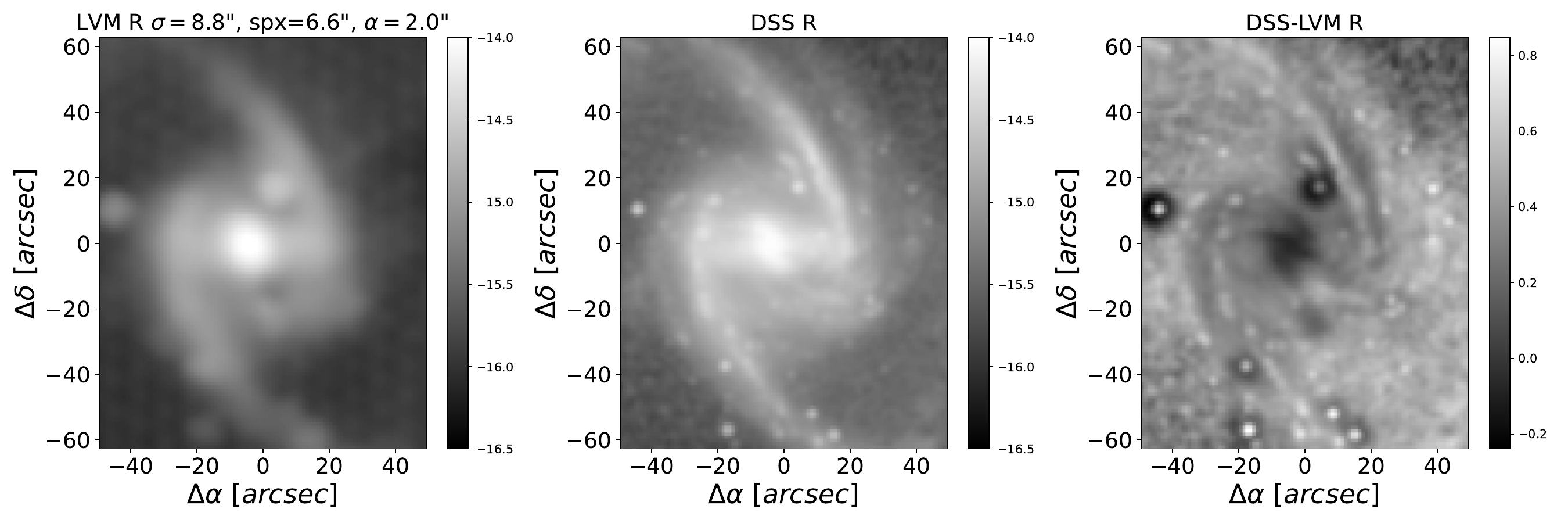}
\includegraphics[width=2\columnwidth]{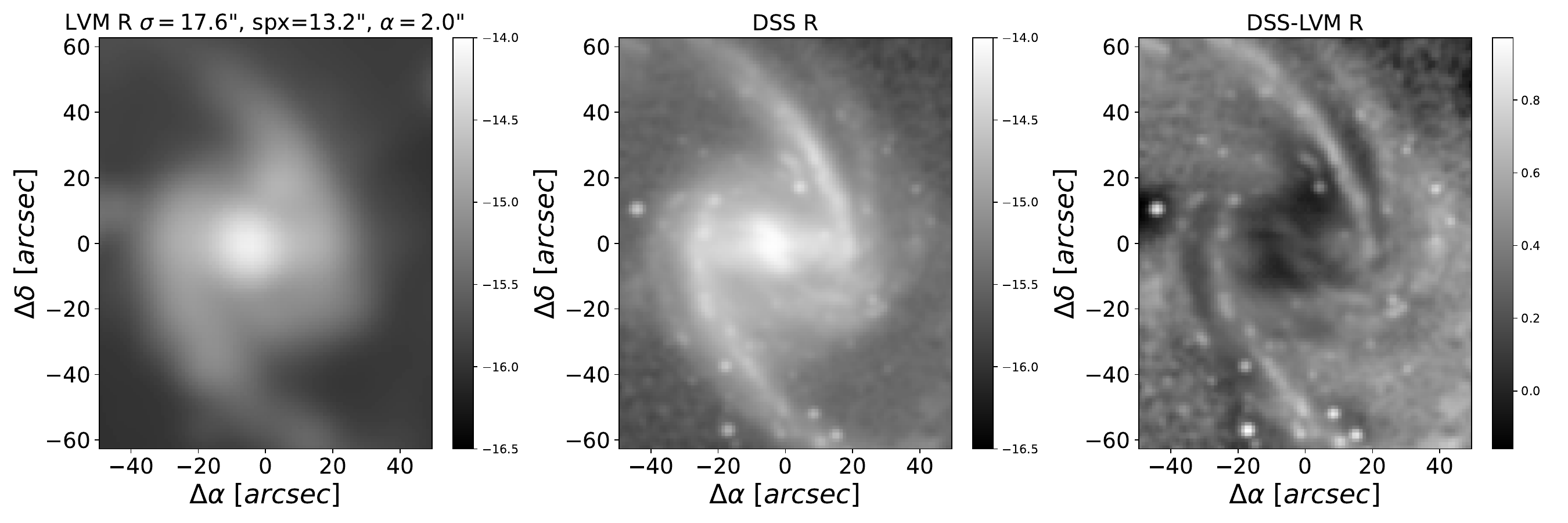}
\caption{Reconstructed images from the 2D interpolation process of NGC 1365. From top to bottom panels: case for $\sigma=4.4''$ ($1/8D_{fib}$, upper panel), case for $\sigma=8.8''$ ($1/4D_{fib}$, middle panel), case for $\sigma=17.6''$ ($1/2D_{fib}$, bottom panel). The value of $\alpha=2$ is constant for all cases. From left to right panels: 2D reconstructed image in the R photometric band using the 9-dithered exposures of the LVM data (left), DSS2 image in the R photometric band of the same region (middle), and the residual image from both images (right). The colour scale and bar are in log $erg/s/cm^2/$\AA\ for all panels.}
\label{fig:recons}
\end{figure*}

The \textsc{3dcubegen} data-cube reconstruction consists of three principal steps: i) The 2D spatial interpolation. ii) The kernel selection, and iii) 3D cube reconstruction. Figure~\ref{fig:workflow} summarises the complete workflow, from the LVM-DRP input spectra to the final data products, including the two optional operating modes described below (steps iv and v). We proceed to describe each step in detail:

\begin{itemize}
    \item[i)] \textbf{The 2D spatial interpolation}: {\sc 3dcubegen} performs a 2D spatial interpolation of the flux $F_{i}(\lambda,x_i,y_i)$ of each fibre located at the sky position $x_i,y_i$ at a given wavelength $\lambda$, into a discrete 2D spatial grid using the next expression:

\begin{equation}
F_{x,y}(\lambda)=\sum_i^{n_{fib}}w(\lambda,r_i)f_AF_{i}(\lambda,x_i,y_i)/\sum_i^{n_{fib}}w(\lambda,r_i).
\label{spatial_rs}
\end{equation}
     Here, $F_{i}(\lambda,x_i,y_i)$ is the $i-th$ spectrum from the RSS lvmSFrame array of each observed dithers and tiles within the data cube FoV. Each spectrum is corrected by the heliocentric velocity shift previous to this step using the individual epochs of each observation.  The factor $f_A$ is the aperture correction between the fibre aperture area $A_{fib}$ and the final spaxel area $A_{spax}$ defined as: $f_A=A_{spax}/A_{fib}$. The value $w(\lambda,r)$ is a weighted kernel defined as $w(\lambda,r)=exp\left(-0.5\left(\frac{r(\lambda)}{\sigma}\right)^{\alpha}\right)$, with $\alpha$ as a sharpener factor of the kernel\footnote{A value of $\alpha=2$ returns a Gaussian kernel shape.}. In addition, $r(\lambda)$ is defined as the distance from the spaxel located at the position ($x,y$) to the centre of the fibre $i$ ($x_i,y_i$) at a given wavelength. It is important to clarifying the distinct roles of the two scale-related quantities in Equation~\ref{spatial_rs}: the aperture correction factor $f_A$ and the kernel size $\sigma$. The factor $f_A = A_\mathrm{spax}/A_\mathrm{fib}$ is a purely photometric correction that rescales the fibre surface brightness to the spaxel area, ensuring flux conservation regardless of the relative sizes of the fibre and the output spaxel. It carries no spatial information and does not affect the weight assigned to any fibre; rather, it represents a unit conversion between the fibre-integrated flux and the per-spaxel surface brightness of the reconstructed cube. On the other hand, the kernel size $\sigma$ controls the spatial resolution of the reconstruction: it determines how steeply the weight $w(r) = \exp(-0.5\,(r/\sigma)^\alpha)$ falls with distance, and therefore how localised the flux interpolation is around each output spaxel. Therefore, Equation~\ref{spatial_rs} is instrument-agnostic: the algorithm treats each fibre as a point source at its astrometric sky position $(x_i, y_i)$, and the physical extent of the fibre aperture enters only through the lower bound on the physically meaningful range of $\sigma$. Specifically, setting $\sigma$ smaller than the fibre diameter $D_\mathrm{fib}$ would produce an oversampled reconstruction in which the inter-fibre gaps of the IFU become visible as artefacts in the output image (see Section~\ref{sec:chi}), because the algorithm attempts to resolve spatial structure at scales where no independent information exists. The fibre diameter therefore acts as a resolution floor: it sets the minimum $\sigma$ below which the reconstruction is no longer physically meaningful, independently of the telescope aperture, the atmospheric seeing, or the observing wavelength. This distinction is the key difference between the LVM and previous IFS surveys such as MaNGA and CALIFA, where the seeing-limited PSF ($\sim\!1$ to $2''$) was the dominant resolution constraint and $D_\mathrm{fib}$ was largely irrelevant. For the LVM, the fibre diameter of $35.3''$ far exceeds the typical LCO seeing ($\sim\!0.8''$), so $D_\mathrm{fib}$ alone sets the physically meaningful range of $\sigma$, as demonstrated by the $\chi^2$ analysis presented in Section~\ref{sec:chi}. Finally, $F_{x,y}$ is the final interpolated flux given at the spaxel at the position $x,y$. Therefore, the total flux at the spaxel ($x,y$) is the average flux at a given wavelength of all the fibres weighted by their distances to that spaxel, i.e. the flux of any given spaxel will be dominated by the flux of the nearest fibre. To minimize the computational time and improve the efficiency of the cube reconstruction, {\sc 3dcubegen} considers only fibres located within a threshold distance from a given spaxel. This distance is two times the kernel size ($2\sigma$). Independent of the value of $\alpha$, the weighted flux contribution of a fibre farther than $2\sigma$ can be depreciated. However, when the kernel size is less than seven times the fibre diameter ($D_{fib}$), the threshold distance is fixed to 7 times the fibre diameter, whatever the kernel size. 
     
     \item[ii)] \textbf{The Kernel selection}: Due to the LVM FWHM point spread function (FWHM-PSF) being $\approx 0.8$\arcsec\footnote{Typical ranges from 0.75\arcsec to 1.25\arcsec}, and the LVM fibre diameter being 35.3\arcsec, the main resolution limit is given by the fibre diameter contrary to MaNGA and CALIFA. Therefore, the safest rule of thumb to fix the minimum kernel size for one-dither observation is to set the kernel diameter to 2 times the fibre diameter ($\sigma=D_{fib}$), which is equivalent to setting the kernel to two times the minimum sampling size. However, due to the LVM dither patterns shown in Figure\ref{fig:dither}, and following the test described in \citet{Sanchez+2023}, it is possible to set the kernel to 8.8\arcsec (1/4$D_{fib}$), with a spaxel size of $0.75\sigma$ as shown in Figure~\ref{fig:recons}. In Section \ref{sec:chi} we describe the best selection of the kernel size without losing spatial information in the 2D interpolation step. 
     %That implies reconstructing 12,400 2D  spatial interpolations.
     \item[iii)] \textbf{3D Cube reconstruction}: {\sc 3dcubegen} repeats the 2D spatial interpolation per each spectral pixel. For the case of the LVM, the spectral range goes from 3600 to 9800 \AA\ with a spectral sampling of 0.5\AA. In addition, it is possible to correct any fibre displacement due to the differential atmospherical refraction (ADR); however, due to the size of the fibre being larger than the maximum refraction shifts, the impact of the ADR is minimal. The final product is a data cube that combines all the spectra from all the tiles observed within a given FoV. This combination produces a perfect way to co-add all the tiles and dithers without losing spatial information and boosting the signal to noise, see Section \ref{sec:decv}.
     
     \item[iv)] \textbf{2D image reconstruction} (optional): As an optional step, {\sc 3dcubegen} can reconstruct 2D maps from the synthetic photometric data ot the RSS spectra or any spectral quantity derived from them like the outputs from the {\sc LVM-DAP} analysis and other RSS analysis. That includes synthetic photometric maps, emission line flux maps, and stellar continuum maps by applying the same kernel 2D interpolation of Equation~\ref{spatial_rs} directly to the data rather than to the full spectral cube. This is equivalent to interpolating any derived parameters from the individual RSS fibres onto a rectified spatial grid, bypassing the need for a full 3D cube reconstruction and the subsequent spectral refitting. This produces spatially registered maps with the corresponding propagated error maps, without requiring the reconstruction of the complete 3D cube. This approach is computationally efficient for exploratory analysis or when only specific spectral diagnostics are needed across a large field or well suited for kinematic analysis at maximum spatial resolution, where working directly from the DAP RSS products preserves the full spectral information of each individual fibre and provides a natural test bed for deconvolution methods applied to resolved stellar and gas kinematics (see Section~\ref{sec:decv} for a discussion of deconvolution in the context of the LVM spatial resolution)

     %\item[iv)] \textbf{2D image reconstruction} (optional): As an optional step, {\sc 3dcubegen} can reconstruct 2D maps of any spectral quantity derived from the \textsc{lvm-dap} analysis --- including synthetic photometric images, emission line flux maps, stellar continuum maps, and kinematic fields such as velocity shifts and velocity dispersions --- by applying the same kernel-weighted 2D interpolation of Equation~\ref{eq:interp} directly to the \textsc{dap} output spectra rather than to the full spectral cube.  
     %This produces spatially registered maps at any of the ten kernel configurations of Table~\ref{tab:kernels}, with the corresponding propagated error maps, at a computational cost orders of magnitude lower than the full cube reconstruction. This approach is particularly well suited for kinematic analysis at maximum spatial resolution, where working directly from the \textsc{dap} RSS products preserves the full spectral information of each individual fibre without the spatial smoothing introduced by the cube reconstruction, and provides a natural test bed for deconvolution methods applied to resolved stellar and gas kinematics (see Section~\ref{sec:deconv} for a discussion of deconvolution in the context of the \textsc{lvm} spatial resolution).

     \item[v)] \textbf{Cube-to-RSS conversion} (optional): As a final optional step, {\sc 3dcubegen} can reproject the reconstructed data cube back into RSS format, producing a new set of spatially co-added spectra suitable for direct ingestion into spectral analysis pipelines such as the {\sc LVM-DAP} (see Section~\ref{sec:dapimp} for details). The error spectra are propagated through the conversion. This step is particularly useful when the 9-dithered coaddition is intended to increase the effective \mbox{S/N} per spectrum before spectral fitting, rather than to produce a spatial image.
\end{itemize}

\begin{figure}[!t]
\centering 
\includegraphics[width=\columnwidth]{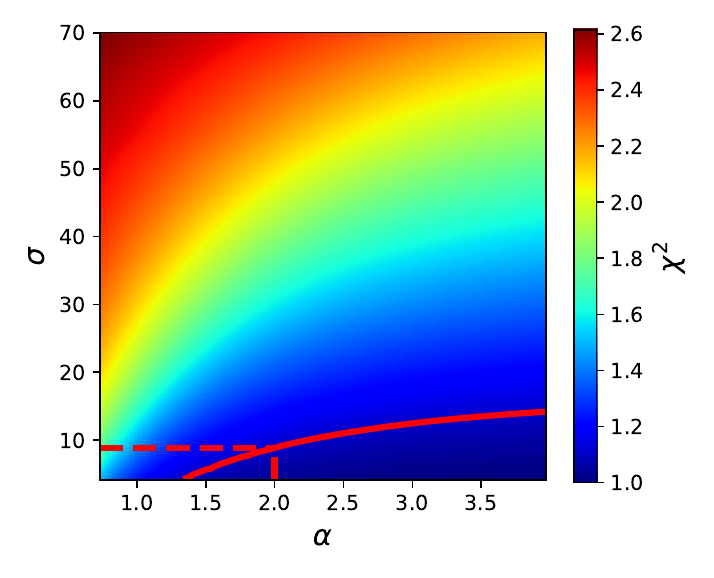}
\caption{$\chi^2$ distribution at multiple kernel values $\alpha$ and $\sigma$. The red solid line represent the limit from an oversampling case. The dashed line marks the values with $\sigma=8.8''$ ($1/4D_{fib}$) and $\alpha=2$ (Gaussian shape). }
\label{fig:Chi2}
\end{figure}

\begin{figure*}[!t]
\centering 
\includegraphics[width=2\columnwidth]{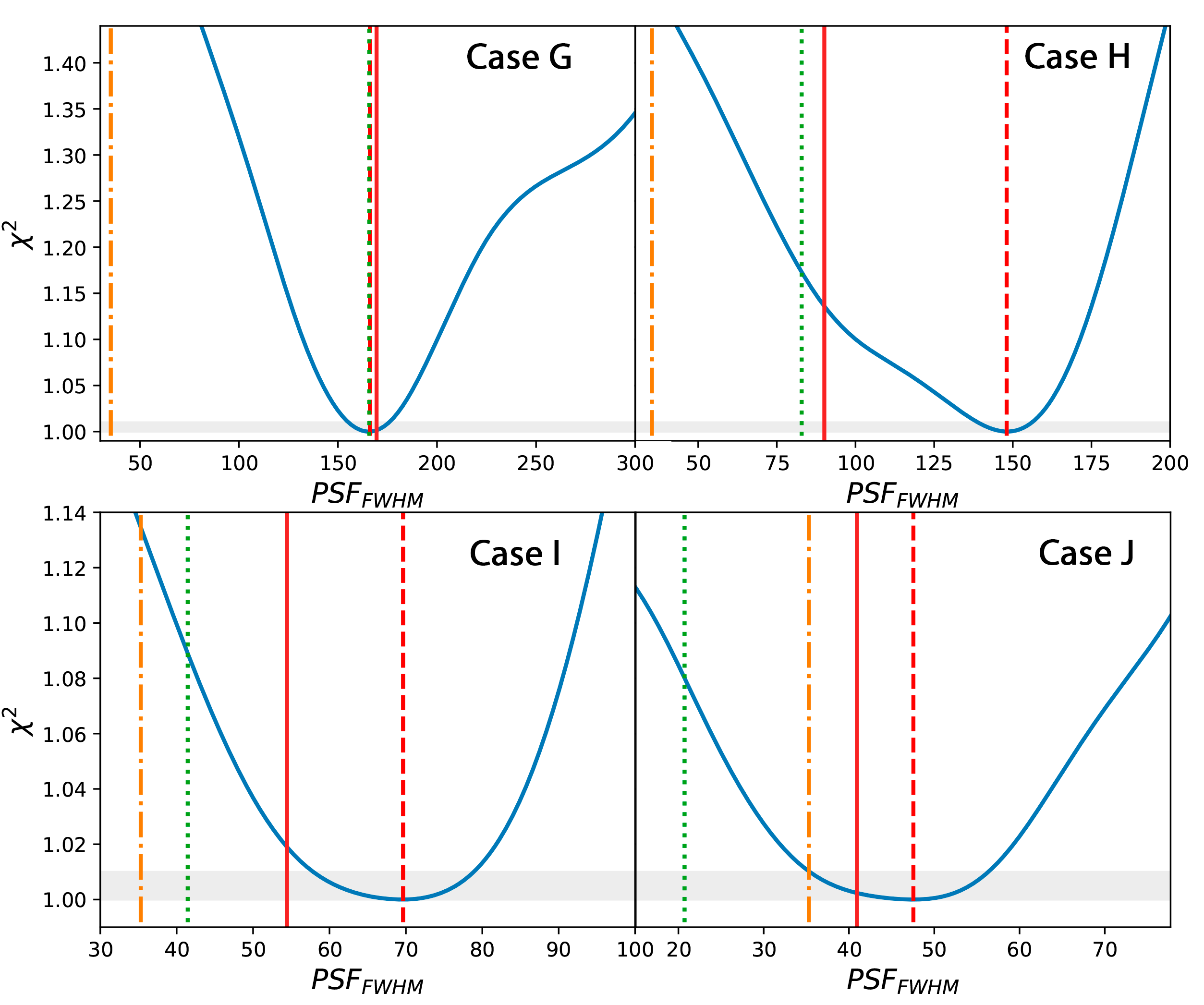}
\caption{Values of the $\chi^2$ values at different deconvolved PSF, with $\alpha=2$ and $\sigma$ equals to $2D_{fib}$ for case G, $D_{fib}$ for Case H, $1/2D_{fib}$ for Case I and $1/4D_{fib}$  for Case J. The red segmented line represents the value at the minimum $\chi^2$, the dotted green line represents the FWHM of the 2D interpolation kernel, the orange segmented-dot line represents the LVM $D_{fib}$, and the solid red line represents the quadratic sum of the FWHM kernel and the $D_{fib}$.}
\label{fig:decX}
\end{figure*}

The error extension of each data cube is computed following the kernel-weighted inverse-variance propagation described in Section~\ref{app:covariance}, where we also  characterise the spatial covariance between neighbouring spaxels introduced by the interpolation and provide practical guidance for downstream spectral extraction.

At the edges of the field of view, the number of fibres contributing to each spaxel decreases and the contribution becomes one-sided. Since Equation~\ref{spatial_rs} is a normalised weighted mean, the reconstruction remains formally defined beyond the outermost fibres; however, in this regime the flux is an extrapolation dominated by the kernel tails rather than by independent spatial information, and the per-spaxel variance increases accordingly (see Section~5.1). To avoid including these unreliable regions in the final products, it is recommended to masks all spaxels whose total accumulated weight $S_N(x,y)$ falls below a threshold fraction of the peak weight of
the field. This criterion naturally traces the hexagonal footprint of the dithered tiles, as can be seen in Figures~\ref{fig:LMC} and \ref{fig:SMC}, and ensures that the edge spaxels retained in the cubes are supported by actual fibre coverage. %For mosaics combining multiple tiles, the same criterion applies at internal tile boundaries, where the overlap between adjacent hexagons compensates the coverage loss.

A related question is whether the kernel interpolation can amplify instrumental or reduction artefacts, such as residuals from an imperfect sky subtraction in individual fibres. Since Equation~\ref{spatial_rs} is a convex weighted mean (the normalised weights sum to unity), the interpolation cannot amplify the signal of any input fibre: the reconstructed flux at any spaxel is bounded by the fluxes of the contributing fibres. The effect of a deviant fibre is instead a spatial redistribution: its residual is spread over the neighbouring spaxels within the kernel support ($r\lesssim 2\sigma$), appearing as a coherent feature at the kernel scale rather than as a single-fibre defect. In the 9-dithered co-addition, such a feature is further diluted by the contributions of the remaining dithers, by a factor set by the relative weights of Equation~\ref{eq:snrgain}. We note, however, that sky-subtraction residuals are not independent noise between fibres of a given exposure, since they originate from a common sky model: systematic residuals shared by all fibres (e.g. around bright airglow lines) are therefore not reduced by the coaddition and will persist in the reconstructed cubes at the same level as in the input RSS. The identification and correction of such systematic sky residuals pertains to the LVM-DRP sky-subtraction module rather than to the cube reconstruction itself.

\section{Chi2 exploration}\label{sec:chi}

Following \citet{Sanchez+2023}, we explore the best reconstruction values for $\alpha$ and $\beta$. We quantify how well 2D interpolation can recover the spatial information of the target. In this exploration, we use observations taken by the LVM for the galaxy NGC1365. NGC1365 has stars, bars, spiral arms, and small structures ideal for quantifying the spatial reconstruction method. We reconstruct the 2D photometric synthetic image from the spectroscopic data of the LVM in the R band (see step {\sc iv} in Section~\ref{sec:cubrec} for details) and compare it with the R band image from the UK Schmidt Telescope taken from the Digitalized Sky Survey\footnote{\url{https://archive.eso.org/dss/dss}} \citep[DSS, see][]{DSS,DSS2a,DSS2b}. 

The DSS images are ideal for this study due to its FWHM-PSF is $\lesssim2''$ \citep{Tritton+1978} that is much lower than the fibre diameter. First, we reconstruct the LVM data, changing $\alpha$ between 0.7 to 4 and $\sigma$ between 4'' to 70'' in steps of 0.01, fixing the spaxel size to 0.75$\sigma$. Then, using an astrometric match, we select a region of 100''$\times$130'' centred in NGC1365 from the LVM and DSS data, making sure to select the same region. Finally, we resample the reconstructed LVM image to match the pixel scale of the DSS data to obtain a residual image between the DSS and the reconstructed image from the LVM data. In Figure \ref{fig:recons} we show an example of this procedure for $\sigma=4.4'',8.8''$ and 17.6'' with $\alpha=2.0$, with the residual defined as $\log_{10}\text{DSS}-\log_{10}\text{LVM}$. To quantify the differences between the 2D interpolation and the DSS image, we use a $\chi^2$ from the next expression:
\begin{equation}
\chi^2 = \sum \left( \log_{10} \left( \frac{\text{DSS}}{\text{LVM}} \right) - Ct \right)^2,
\label{eq:chi}
\end{equation}
where $Ct=0.61$ is a normalization constant between the photometric offset of the DSS photometry and the LVM data, with the summation performed over all pixels within the region.

\begin{figure*}[!t]
\centering 
\includegraphics[width=2\columnwidth]{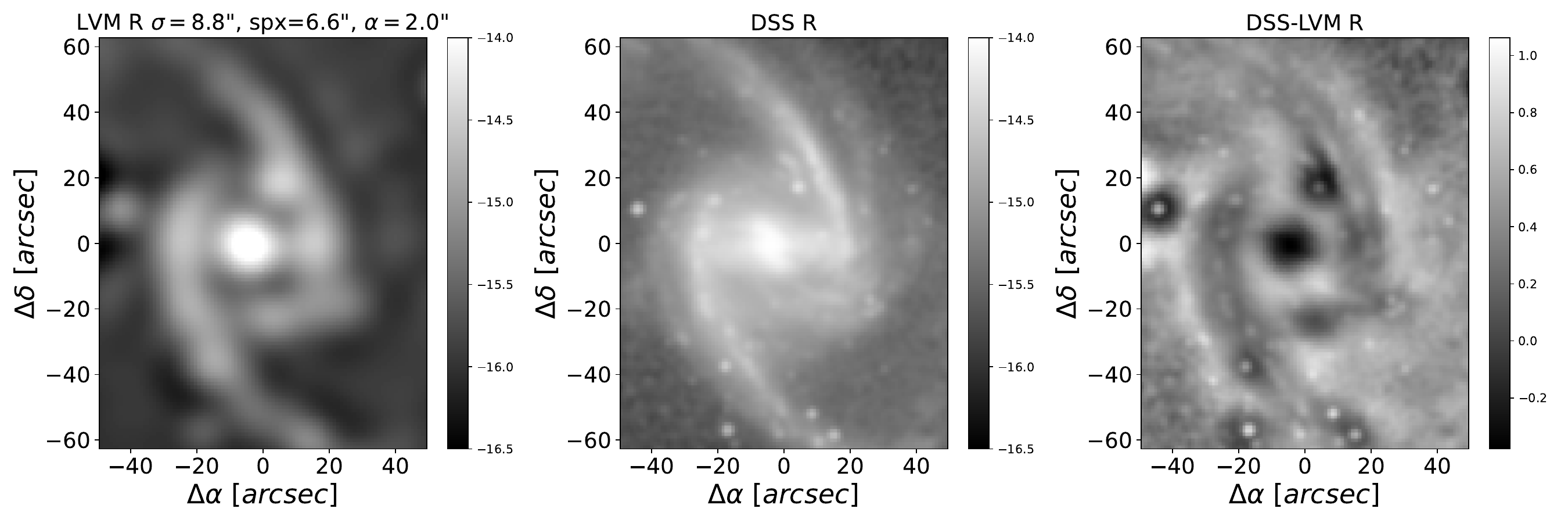}
\includegraphics[width=2\columnwidth]{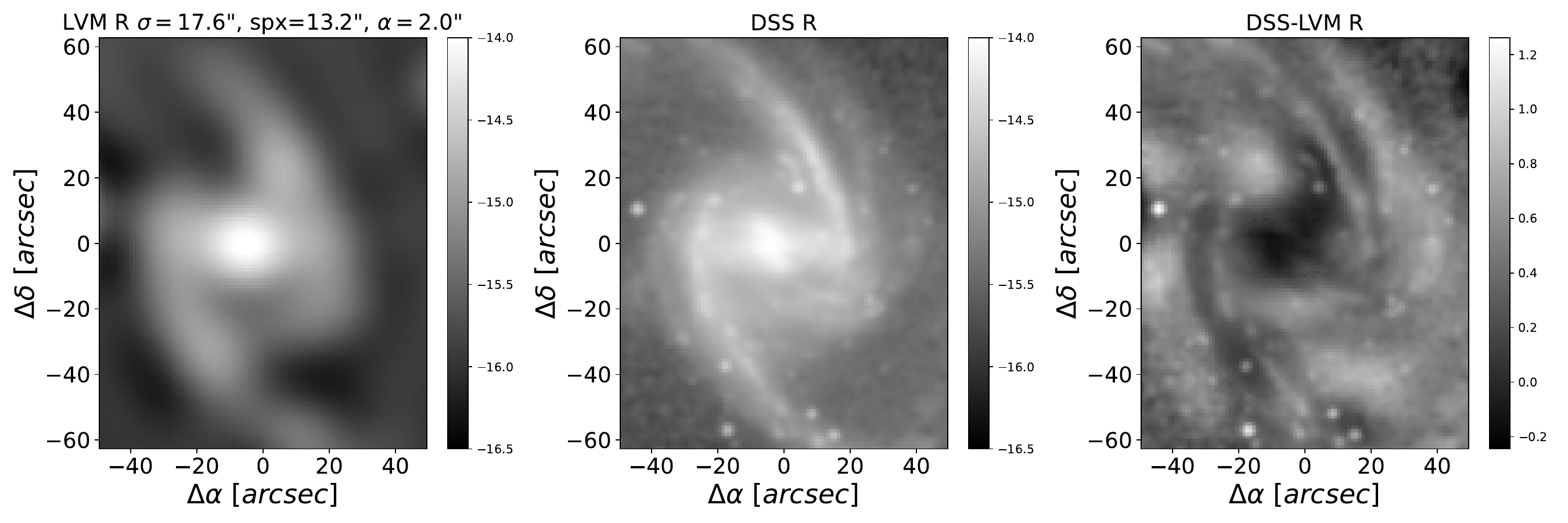}
\caption{Deconvolved reconstructed images of NGC 1365 using the best fitted PSF values. Upper panel, case for $\sigma=8.8''$ ($1/4D_{fib}$ and Case J), with a deconvolved FWHM PSF of $47.5''$. Lower panel, case for $\sigma=17.6''$ ($1/2D_{fib}$ and Case H), with a deconvolved FWHM PSF of $69.6''$. Left panel: 2D reconstructed image in the R photometric band using the 9-dithered exposures of the LVM data. Middle panel: DSS2 image in the R photometric band of the same region. Right panel: residual image. The colour scale and bar are in log $erg/s/cm^2/$\AA\ for all panels.}
\label{fig:dec}
\end{figure*}

The final results are shown in Figure \ref{fig:Chi2}. We find that $\chi^2$ decreases for smaller values of $\sigma$ and larger values of $\alpha$. This result is a consequence of the nature of both parameters: larger $\alpha$ implies a sharp kernel shape, and a small $\sigma$ returns a small kernel size. Therefore, there is a degeneracy between $\alpha$ and $\sigma$ that returns equivalent kernel values in size and shape. 

It is important to clarifying the choice of reference image in this comparison. Rather than comparing the reconstructed LVM image against a version of the DSS convolved with the LVM fibre aperture (which would only quantify the residuals attributable to the interpolation kernel) we compare against the unconvolved DSS image intentionally. This choice allows us to quantify the total loss of spatial information introduced by the combined effect of the 35.3$''$ circular fibre aperture and the interpolation kernel $w(r)$, which is the physically meaningful quantity for assessing the spatial fidelity of the reconstructed cubes. In other words, the $\chi^2$ defined in Equation~\ref{eq:chi} measures how well the reconstruction recovers the intrinsic sky brightness distribution, not merely how well it reproduces what the fibres could have measured in the first place. This is a deliberately conservative metric: the minimum achievable $\chi^2$ is set by the information lost to the fibre aperture alone, independently of the reconstruction method, and corresponds to the limiting case where the DSS image is convolved with the 35.3$''$ circular top-hat aperture. %The difference between this floor and the measured $\chi^2$ at any given $(\sigma, \alpha)$ therefore quantifies the additional spatial degradation introduced by the interpolation kernel, providing a clean separation between aperture-limited and reconstruction-limited contributions to the spatial resolution budget.

However, a lower $\chi^2$ value does not always indicate a better 2D reconstruction. As it is shown in Figure \ref{fig:recons}, when the kernel size ($\sigma=4.4''$) is too small in comparison with the fibre size $D_{fib}$, the patterns of the dithering are visible, showing an oversampling case. However, the residuals tend to reach an average constant value because, due to the oversampling, the spatial resolution is limited by the combination of the dither pattern and the fibre size: it is not possible to recover smaller structures. In addition, since the PSF at LCO ($\sim0.8''$) is much smaller than the LVM $D_{fib}$ ($35.2''$), the $\chi^2$ tends to be smaller as larger $\alpha$ and lower $\sigma$ is but reaching steadily into a stable value. 

Therefore, we select a limit where the kernel size has a value of $\sigma=8.8''$ and $\alpha=2$, thus, when the kernel in a Gaussian shape has a diameter of half of the diameter of the fibre $D_{fib}$. It ensures no oversampling of the 2D interpolation and recovers any possible structure limited by $D_{fib}$ ($\sigma=1/4D_{fib}$). This limit also implies any combination of $\sigma$ and $\alpha$ that recovers the exact value of $\chi^2$ as recomended by \citet{Sanchez+2023}. In Figure \ref{fig:Chi2}, we over-plotted all the possible combinations of $\sigma$ and $\alpha$ that mark the limit of the oversampling case. Any point below the continuous red line will be oversampled with the resolved spatial structures limited by $D_{fib}$. Therefore, at this point, to recover any real structures, it will be necessary to deconvolve the reconstructed image from the actual PSF of the image \citep{Sanchez+2023}.

\subsection{PSF Deconvolution}\label{sec:decv}

To explore the previous point, we proceed to implement the Richardson-Lucy algorithm deconvolution \citep{Richardson+72} from the {\sc RESTORATION} module of {\sc scikit-image}\footnote{\url{https://scikit-image.org/}} package \citep{vanderWalt2014}. We fix $\alpha=0$, and define four cases with $\sigma=8.8''$, 17.6'', 35.2'' and 1.2'' (1/4, 1/2, 1 and 2 times the $D_{fib}$), hereafter cases J, I, H and G. We use a Gaussian profile to model the deconvolved PSF with a dispersion that varies from 0.3 to 5 pixels in steps of 0.01 pixels, fixing the spaxel size to 0.75$\sigma$ for each case. Then, we deconvolve the reconstructed LVM image with the PSF model and quantify its residuals using Equation \ref{eq:chi}. In Figure \ref{fig:decX}, we show the results of $\chi^2$ at different values of the FWHM of the deconvolved PSF\footnote{Defined as $2\sqrt{2\ln2}$ of the PSF dispersion}, here after $PSF_{FWHM}$. 

We find minimum $\chi^2$ values when the $PSF_{\rm FWHM}$ has 47.5'', 69.6'', 148.1'', and 166.1'' (corresponding to 3.1, 2.2, 2.4, and 1.3 pixels) for the J, I, H, and G cases, respectively, as indicated by the segmented red line in Figure \ref{fig:decX}. These values represent the effective PSF sizes of the reconstructed images resulting from the 2D interpolation process. To better understand the origin of these effective PSF values, we compare them with both the FWHM of the 2D interpolation kernel ($2\sqrt{2\ln2}\sigma$, shown as the dotted green line) and the fibre diameter $D_{\rm fib}$ (segmented-dotted orange line). We find that the relative contribution of the fibre size and interpolation kernel varies between cases. For case G, the spatial resolution is primarily determined by the interpolation kernel, indicating that the kernel $\sigma$ dominates the effective PSF. In contrast, for cases H and I, the fibre size becomes increasingly important. In these cases, the quadratic combination of the fibre diameter and kernel size, $\sqrt{D_{\rm fib}^2 + 8\ln2\sigma^2}$ (solid red line), closely reproduces the effective PSF of the reconstructed images. For case H, the kernel still plays a dominant role, and the quadratic combination of both components marks the point where $\chi^2$ reaches its minimum valley. For case I, the fibre size and kernel contribute equally, and again the quadratic combination traces the onset of the $\chi^2$ valley. Finally, in case J, the fibre diameter fully dominates the effective PSF, with the kernel contributing only marginally. Therefore, for case J and small kernel sizes, the minimum achievable spatial resolution is determined primarily by the fibre size rather than by the interpolation kernel.

To obtain spatial information for smaller values of the fibre diameter, the deconvolution process could help. In Figure~\ref{fig:dec}, we show the deconvolution results for the cases J and I using their effective PSF values for the deconvolution. While mathematically, the deconvolution can improve the spatial resolution of an image, from the visual inspection of Figure~\ref{fig:dec}, it is clear that this process create spurious structures that are not real and, therefore, provide false spatial information of the object. Those spurious structures are clearly shown in the upper of  Figure~\ref{fig:dec} that represent the J case: the shape of the bulge of NGC1365 is not well recovered as it recovered for its non-deconvolved version (middle panel of Figure \ref{fig:recons}). For case I, the deconvolved image is similar in resolution to case J; therefore, deconvolution is unnecessary. Therefore, deconvolution is not recommended when attempting to recover spatial structures below the effective fibre aperture of the LVM, as the spatial information at these scales is not present in the data.

We note that the Gaussian profile used here to model the deconvolution kernel is an approximation to the true effective PSF of the reconstructed image. As pointed out by \citet{vanderMarel1995} and \citet{vanderMarel1997}, the physically correct convolution kernel that describes the LVM observation is the convolution of the 35.3$''$ circular fibre top-hat aperture with the interpolation kernel $w(r)$, which for the Gaussian case ($\alpha = 2$) can be evaluated semi-analytically. However, the goal of the deconvolution exercise presented here is not to recover the intrinsic sky brightness distribution (which would require that physically motivated kernel) but rather to empirically measure the effective PSF of the reconstructed image by finding the Gaussian profile whose removal minimises the residuals with respect to the unconvolved DSS reference. In this context, the fitted Gaussian PSF is an effective, empirical quantity that absorbs both the fibre aperture and kernel contributions to the spatial resolution, consistently with the $\chi^2$ metric defined in Section~\ref{sec:chi}. Deconvolution using the physically correct composite kernel, with the goal of recovering spatial information below the fibre aperture, is a more ambitious objective that falls outside the scope of this work and is left for future investigation. The conclusion that deconvolution with an approximate Gaussian kernel generates spurious structures therefore reflects the limitations of that specific approach, not a fundamental limitation of deconvolution in general.

\section{Noise properties of the reconstructed cubes}\label{secc:noise}

\subsection{Error propagation and spatial covariance}
\label{app:covariance}

\begin{figure}
    \centering
    \includegraphics[width=0.85\columnwidth]{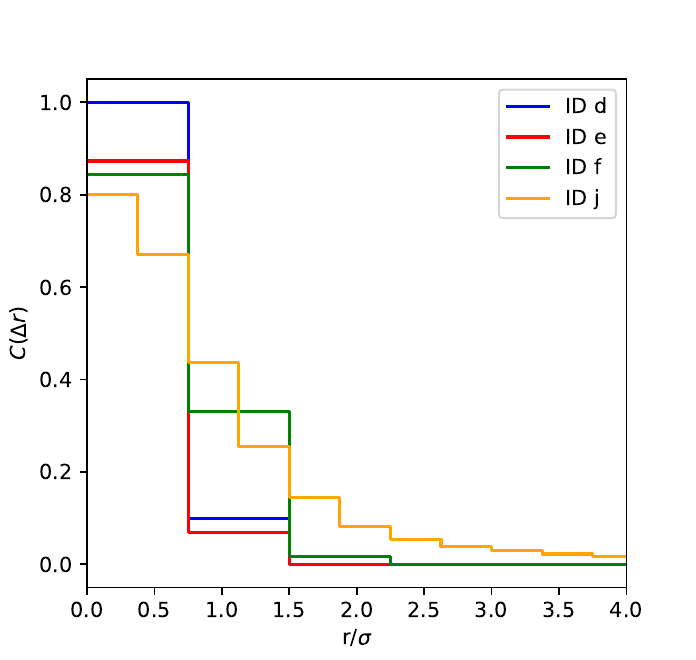}
    \caption{Normalised 2-point correlation function $C(\Delta r)$ of the reconstructed cube noise as a function of spaxel separation in units of the kernel size $\sigma$, computed from \textsc{lvm} observations of N44 for kernel configurations $f$ (blue), $g$ (red), $h$ (green), and $i$ (orange).}
    \label{fig:app:covariance}
\end{figure}

\begin{figure*}
    \centering
    \includegraphics[width=\textwidth]{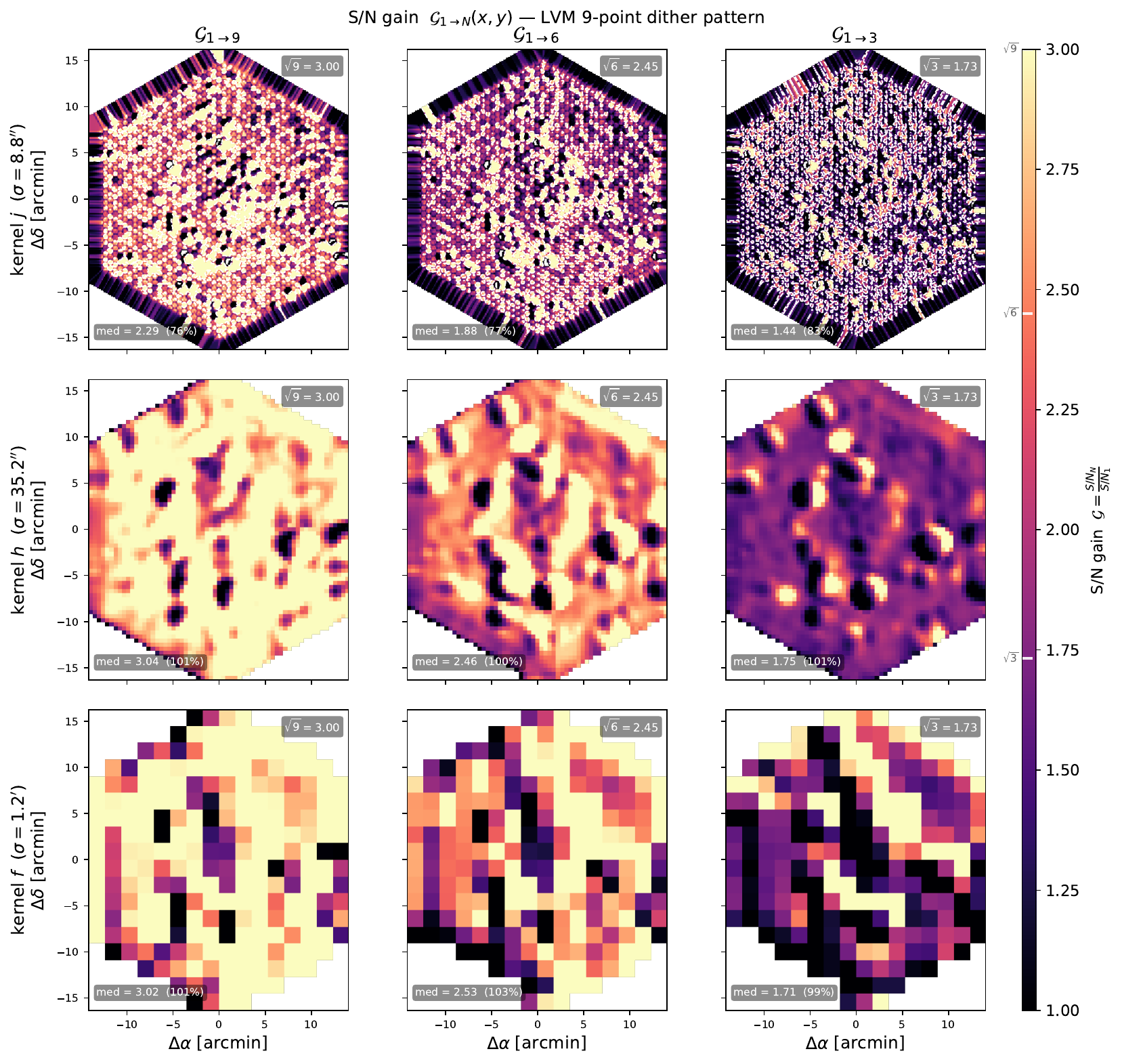}
    \caption{Spatial distribution of the \mbox{S/N} gain $\mathcal{G}_{1\to N}(x,y) = \mathrm{S/N}_N(x,y)\,/\,\mathrm{S/N}_1(x,y)$ measured directly from the reconstructed weight maps for LVM observations of N44. Columns show the gain from combining 9, 6, and 3 dithers (left to right); rows show kernel configurations $j$ ($\sigma = 8.8''$), $h$ ($\sigma = 35.2''$), and $f$ ($\sigma = 1.2'$) from top to bottom. The colour scale is common to all panels and spans $\mathcal{G} = 1$ to $\sqrt{9} = 3$; the dashed contour in each panel marks the theoretical $\sqrt{N}$ expectation. Black regions correspond to spaxels with fewer than 0.5\% of the peak weight and are masked.}
    \label{fig:snrmap}
\end{figure*}

The 2D spatial interpolation described in Equation~\ref{spatial_rs} can be written compactly in matrix form as

\begin{equation}
    \mathbf{F}^{N,M} = \mathbf{W}^{N,M,K}\,\mathbf{F}^{K},
    \label{eq:app:matrixform}
\end{equation}

where $\mathbf{F}^{N,M}$ is the reconstructed flux cube indexed over the $N \times M$ spatial grid, $\mathbf{F}^K$ is the vector of $K = n_\mathrm{fib}$ input RSS spectra, and $\mathbf{W}^{N,M,K}$ is the normalised weight tensor whose elements are $W_{i,j,k} = w(r_{i,j,k})/\sum_k w(r_{i,j,k})$, with $w(\lambda, r) = \exp\!\left(-0.5\,(r/\sigma)^\alpha\right)$ as defined in Equation~\ref{spatial_rs}. The RSS fibre errors are assumed to be uncorrelated between fibres: a valid approximation since the \textsc{lvm-drp} propagates independent detector noise, so the fibre error covariance matrix $\epsilon^{K,K}$ is diagonal:

\begin{equation}
    \epsilon^{K,K} = \mathrm{diag}\!\left(
        \sigma^2_{1,1},\; \sigma^2_{2,2},\; \ldots,\;
        \sigma^2_{n_\mathrm{fib},\,n_\mathrm{fib}}
    \right).
    \label{eq:app:fibcov}
\end{equation}

The full spatial covariance tensor of the reconstructed cube then follows from standard error propagation:

\begin{equation}
    \boldsymbol{\varepsilon}^{N,M,N,M} =
        \mathbf{W}^{N,M,K}\;\epsilon^{K,K}\;\mathbf{W}^{K,N,M},
    \label{eq:app:covtensor}
\end{equation}

whose elements are

\begin{equation}
    \varepsilon_{i,j,l,m} =
        \sum_{k_1=1}^{n_\mathrm{fib}}\sum_{k_2=1}^{n_\mathrm{fib}}
        W_{i,j,k_1}\;\sigma^2_{k_1,k_2}\;W_{l,m,k_2}.
    \label{eq:app:covelem}
\end{equation}
Because $\epsilon^{K,K}$ is diagonal ($\sigma^2_{k_1,k_2} = \sigma^2_k\,\delta_{k_1 k_2}$), Equation~\ref{eq:app:covelem} reduces to
\begin{equation}
    \varepsilon_{i,j,l,m} = \sum_{k=1}^{n_\mathrm{fib}}
        W_{i,j,k}\;\sigma^2_k\;W_{l,m,k}.
    \label{eq:app:covreduced}
\end{equation}
This expression has two physically distinct cases. In \textbf{Case a} ($i=l$, $j=m$), the diagonal elements $\varepsilon^{N,M}$ yield the per-spaxel error map stored in the second \textsc{fits} extension of each data cube --- the standard propagated uncertainty accessible to all users. In \textbf{Case b} ($i,j \neq l,m$), the off-diagonal elements encode the spatial covariance between distinct spaxels at separation $r$, which defines the 2-point covariance function of the cube.

Because the same input fibres contribute to multiple neighbouring spaxels within the kernel support radius $2\sigma$, the weight vectors $W_{i,j,k}$ and $W_{l,m,k}$ overlap for spaxels separated by less than $\sim\!2\sigma$, making $\varepsilon_{i,j,l,m} > 0$. The per-spaxel error extension (Case a) correctly captures the formal uncertainty at each spaxel in isolation, but does not encode this inter-spaxel covariance. To characterise it quantitatively, \textsc{3dcubegen} includes a dedicated covariance analysis module that computes the full $\boldsymbol{\varepsilon}^{N,M,N,M}$ tensor and its normalised 2-point correlation function

\begin{equation}
    C(\Delta r) = \frac{\varepsilon_{i,j,l,m}}{\varepsilon_{i,j,i,j}},
    \label{eq:app:twopt}
\end{equation}

as a function of spaxel separation $\Delta r$. We applied this analysis to \textsc{lvm} observations of N44 (LMC) for kernel configurations $f$, $g$, $h$, and $i$ (see Table~\ref{tab:resolutions}). The resulting 2-point correlation functions, normalised by their zero-lag values and plotted against $r/\sigma$, are shown in Figure~\ref{fig:app:covariance}. All four configurations follow a self-similar profile: the correlation drops sharply to zero within $r \lesssim 2\sigma$, confirming that the covariance is spatially compact and scales with the kernel size. Spaxels separated by more than $\sim\!2\sigma$ are effectively uncorrelated, regardless of the kernel configuration chosen.

%All four configurations show a self-similar profile that drops to zero within $r \lesssim 2\sigma$, confirming that the inter-spaxel covariance introduced by the kernel interpolation is spatially compact and scales predictably with $\sigma$. Spaxels separated by more than $\sim\!2\sigma$ are effectively uncorrelated.

This spatial covariance has a direct practical consequence for downstream spectral analysis. Users who co-add $N_\mathrm{sp}$ spaxels within an aperture and propagate the stored per-spaxel variances in quadrature implicitly assume that all spaxel errors are independent, which overestimates the true signal-to-noise ratio of the combined spectrum. The ratio of effectively dependent samples within an aperture of radius $r_\mathrm{ap}$ is approximately

\begin{equation}
    R_\mathrm{dep} =\left(\frac{N_\mathrm{sp}}{N_\mathrm{dep}}\right) =\left(\frac{r_\mathrm{ap}}{\sigma}\right)^{\!2},
    \label{eq:app:neff}
\end{equation} with a total spaxel count defined as $N_\mathrm{sp} = \pi(r_\mathrm{ap}/p)^2$, where $p$ is the spaxel size. This gives a \mbox{S/N} overestimation factor of $\sqrt{N_\mathrm{sp}/N_\mathrm{ind}} \approx r_{ap}/\sqrt{r_{ap}^2-\sigma^2}$, with $N_\mathrm{ind}=N_\mathrm{sp}(1-1/R_{dep})$ as the number of independent spaxels for any kernel configuration. We therefore recommend that any aperture or bin level spectral extractions from the cubes need to be accompanied by an empirical noise rescaling factor following the approaches of \citet{Sanchez+2023} and \citet{Law+2016}. For users requiring the complete noise structure for advanced analyses, such as full-spectrum fitting or Voronoi tessellation with correct error weighting \citep[e.g.,][]{Cappellari+2003}, \textsc{3dcubegen} can optionally save the full spatial covariance tensor $\boldsymbol{\varepsilon}^{N,M,N,M}$ as an additional \textsc{fits} extension via the \texttt{fcovmat=True} option, providing a rigorous characterisation of the inter-spaxel noise correlations introduced by the reconstruction.

\subsection{Signal-to-noise improvement from dither coaddition}
\label{sec:snrgain}

%For kernels $h$ and $f$, the median gain reaches $\sim\!100$\% of $\sqrt{N}$ across the field, confirming near-ideal coaddition efficiency when $\sigma \gtrsim D_\mathrm{fib}$. For kernel $j$, the median gain drops to $\sim\!77$\% of $\sqrt{N}$ and the gain map shows strong spatial variation that directly traces the discrete fibre positions of the dither pattern, reflecting the oversampling regime at this kernel size. Black regions correspond to spaxels with fewer than 5\% of the peak weight and are masked.
One of the primary scientific motivations for the 9-point dither strategy is the improvement in signal-to-noise ratio achieved by combining multiple exposures. For a simple coadd of $N$ independent, identically-noisy exposures, the expected \mbox{S/N} improvement is $\sqrt{N}$, giving a factor of 3 for 9 dithers. However, in the \textsc{3dcubegen} kernel interpolation framework, each spaxel flux is a normalised weighted mean (Equation~\ref{spatial_rs}), so the variance at spaxel $(x,y)$ is (Appendix~\ref{app:covariance}, Case a):
\begin{equation}
\begin{aligned}
    \varepsilon(x,y) = \frac{Q_N(x,y)}{S_N^2(x,y)}\,\sigma^2,\\
    \qquad
    S_N = \sum_{d=1}^{N}\sum_{k=1}^{n_\mathrm{fib}} w(r_{d,k}),\\
    \qquad
    Q_N = \sum_{d=1}^{N}\sum_{k=1}^{n_\mathrm{fib}} w^2(r_{d,k}),
    \label{eq:variance}
\end{aligned}
\end{equation}
where $r_{d,k}$ is the distance from the spaxel to the $k$-th fibre in dither $d$, and $\sigma^2$ is the fibre noise variance assumed uniform across all dithers. The \mbox{S/N} at a given spaxel is therefore proportional to $S_N/\sqrt{Q_N}$, and the gain from combining $N$ dithers relative to a single exposure is:
\begin{equation}
    \mathcal{G}_{1\to N}(x,y) 
    = \frac{S_N}{S_1}\sqrt{\frac{Q_1}{Q_N}}.
    \label{eq:snrgain}
\end{equation}
Note that this reduces to $\sqrt{N}$ only when all $N$ dithers contribute with identical weights to every spaxel, i.e. when $\sigma \gg \Delta_\mathrm{dither}$.

For kernel $j$ ($\sigma = 8.8''$, $\alpha = 2$), the dither offsets are comparable to $\sigma$, so the weights vary strongly between dithers. For a spaxel at the field centre, the nearest fibre in dither 1 lies at $r \approx 0''$ ($w \approx 1.00$), the nearest fibres in dithers 4--9 lie at $r \approx 12.3''$ ($w \approx 0.37$), and the nearest fibres in dithers 2--3 lie at $r \approx 21.4''$ ($w \approx 0.06$). 
This gives:
\begin{align}
    S_9 &= 1.00 + 6\times0.37 + 2\times0.06 = 3.34, \notag\\
    Q_9 &= 1.00 + 6\times0.37^2 + 2\times0.06^2 = 1.83, \notag\\
    \mathcal{G}_{1\to9} &= \frac{3.34}{1.00}\sqrt{\frac{1.00}{1.83}} \approx 2.47,
    \label{eq:gain_j}
\end{align}
below the $\sqrt{9} = 3.0$ expectation. For larger kernels where $\sigma \gg \Delta_\mathrm{dither}$, all dithers contribute with $w \approx 1$ and $\mathcal{G}_{1\to N} \to \sqrt{N}$.

%, but also significantly higher than the $\approx\!1.8$ one would estimate from the linear weight sum alone. The difference between these two estimates reflects the effect of normalisation: because each spaxel flux is a weighted \emph{mean} rather than a weighted sum, dithers with low $w$ contribute little to the variance even as they modestly increase $S_N$.

The spatial distribution of $\mathcal{G}_{1\to N}(x,y)$, measured directly as the ratio of the reconstructed \mbox{S/N} maps $\mathrm{S/N}_N /\mathrm{S/N}_1$, is shown in Figure~\ref{fig:snrmap} for kernel configurations $j$, $h$, and $f$ and for $N = 3$, 6, and 9 dithers. The results reveal a clear dependence on kernel size. For kernel $h$ ($\sigma = 35.2''$) and kernel $f$ ($\sigma = 140.8''$), the median gain reaches essentially the theoretical $\sqrt{N}$ expectation across the full field: the median values are $\mathcal{G}_{1\to9} = 3.04$ (101\%), $\mathcal{G}_{1\to6} = 2.46$ (100\%), and $\mathcal{G}_{1\to3} = 1.75$ (101\%) for kernel $h$, and $\mathcal{G}_{1\to9} = 3.02$ (101\%), $\mathcal{G}_{1\to6} = 2.53$ (103\%), and $\mathcal{G}_{1\to3} = 1.71$ (99\%) for kernel $f$. This confirms that when $\sigma \gtrsim D_\mathrm{fib}$, the kernel is wide enough to average over the dither offsets and the coaddition behaves as an ideal independent-exposure stack.

For kernel $j$ ($\sigma = 8.8''$), the behaviour is qualitatively different. The median gains are $\mathcal{G}_{1\to9} = 2.29$ (77\%), $\mathcal{G}_{1\to6} = 1.88$ (77\%), and $\mathcal{G}_{1\to3} = 1.44$ (83\%), all significantly below $\sqrt{N}$. This is a direct consequence of $\sigma$ being smaller than the dither offsets: the kernel weight $w(r) = \exp(-0.5\,(r/\sigma)^2)$ falls steeply with distance, so dithers offset by $\sim\!10$--$20''$ contribute little weight and therefore little noise reduction. The gain map for kernel $j$ also shows strong spatial variation - the alternating bright and dark pattern visible in the top row of Figure~\ref{fig:snrmap} reflects the discrete fibre positions of the dither pattern itself: spaxels that happen to lie near the centre of a fibre in multiple dithers achieve high gain, while those falling in inter-fibre gaps gain little. This is precisely the oversampling regime discussed in Section~\ref{sec:chi}: the spatial structure of the gain map at kernel $j$ is itself a diagnostic of the dither geometry, and spatially averaged quantities (such as the median gain) should be interpreted with this variation in mind.

In addition, from Figure~\ref{fig:snrmap}, we found that the fact that kernels $h$ and $f$ consistently reach $\sim\!100$\% of $\sqrt{N}$ across the full field is a non-trivial validation of the \textsc{3dcubegen} coaddition implementation: any systematic error in the kernel normalisation or double-counting of fibres would produce a measurable deviation from this expectation, and none is observed. 

Second, the strongly structured gain map for kernel $j$ is not noise but a direct imprint of the dither geometry on the reconstruction: spaxels that happen to coincide with fibre centres across multiple dithers achieve high gain, while those falling in inter-fibre gaps gain little, producing the alternating bright-dark pattern visible in the top row. This spatial structure is itself a diagnostic of the oversampling regime discussed in Section~\ref{sec:chi}, and users working at kernel $j$ resolution should be aware that the local \mbox{S/N} can vary substantially across a single resolution element. 

\begin{table}[!t]
    \caption{Values of the 10 kernel configurations to reconstruct the datacubes from the 9-dithered LVM observations of the LV and MC. The spatial resolution in parsec is calculated at the distance of the LMC and SMC and is defined as the physical size of the PSF dispersion.}
    \label{tab:resolutions}
    \centering
    \begin{tabular}{lccccc}
         \toprule
         ID & kernel & spaxel & FWHM & \multicolumn{2}{c}{Spatial Resolution}\\
         & size & size & PSF & \multicolumn{2}{c}{physical}\\ 
         &  &  &  & LMC & SMC \\ 
         \midrule
         a & 1.25$^{\circ}$ & 56.3'  &   2.9$^{\circ}$ &  1.09 kpc &  1.36 kpc \\
         b & 37.6'          & 28.2'  &   1.5$^{\circ}$ & 545.77 pc & 681.96 pc \\
         c & 18.8'          & 14.1'  &  44.3'          & 272.88 pc & 340.98 pc \\
         d & 9.4'           & 7.1'   &  22.1'          & 136.44 pc & 170.49 pc \\
         e & 4.7'           & 3.5'   &  11.0'          &  68.22 pc &  85.25 pc \\
         f & 2.3'           & 1.7'   &   5.4'          &  34.11 pc &  42.62 pc \\
         g & 1.2'           & 54.0'' &   2.8'          &  17.05 pc &  21.31 pc \\
         \hline
         h & 35.2''         & 26.4'' &   2.4'          &  15.22 pc & 19.08 pc \\ %8.53 pc &  10.66 pc \\
         i & 17.6''         & 13.2'' &  69.6''         &   7.16 pc &  8.08 pc \\%4.26 pc &   5.33 pc \\
         j &  8.8''         &  6.6'' &  47.5''         &   4.88 pc &  6.12 pc \\%2.13 pc &   2.67 pc \\
         \bottomrule
    \end{tabular}
\end{table}

\begin{figure*}[!t]
\centering 
\includegraphics[width=\textwidth]{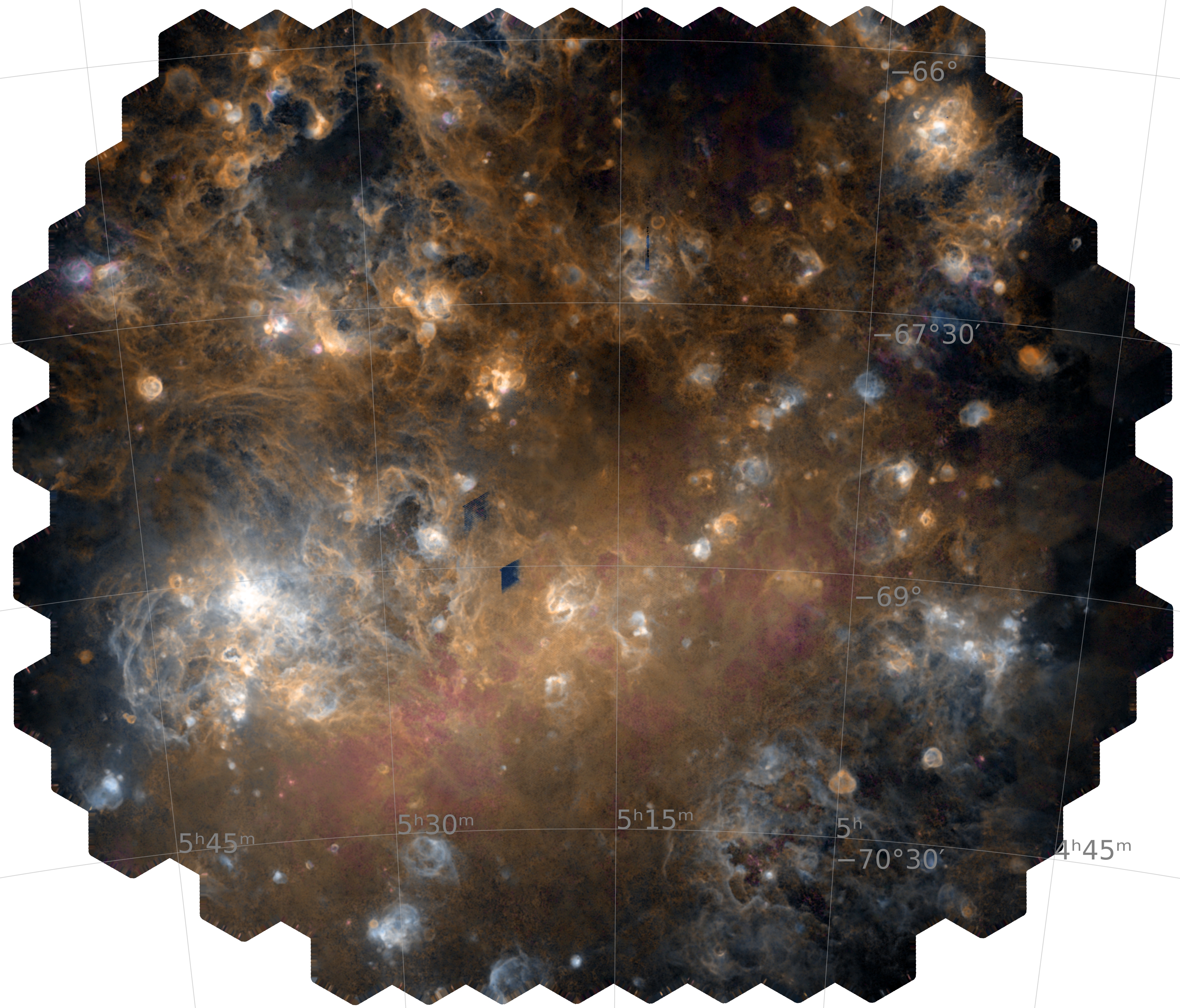}
\caption{Reconstructed colour narrowband image of the full tiled LMC observations using the kernel $j$ configuration ($\sigma=8.8''$, $\alpha=2$). The
RGB composite follows the Hubble palette: \oiii as Blue, \ha as Green, and \sii as red. The H$\alpha$ channel is partially desaturated to enhance the contrast between oxygen- and sulfur-dominated structures, highlighting the complex ionization morphology across the galaxy.}
\label{fig:LMC}
\end{figure*}

Third, the slightly lower median gain for kernel $j$ with 9 dithers (77\%) compared to 3 dithers (83\%) is not a contradiction: adding more dithers with large offsets increases both $S_N$ and $Q_N$, and because the gain scales as $\mathcal{G} \propto S_N / \sqrt{Q_N}$, the marginal contribution of distant dithers to the variance reduction is smaller than their contribution to the weight sum, compressing the percentage relative to $\sqrt{N}$ as $N$ grows. Taken together, these results establish that the \mbox{S/N} improvement from the LVM 9-point dither pattern is not a single number but a kernel-dependent, spatially varying quantity: kernels $h$ and larger deliver near-ideal coaddition efficiency at the cost of spatial resolution, while kernel $j$ preserves maximum spatial resolution at the price of a $\sim\!23$\% deficit in \mbox{S/N} depth relative to the $\sqrt{9}$ expectation

%These results confirm that the \mbox{S/N} improvement from the 9-point dither pattern is not a fixed number but a spatially varying quantity that depends critically on the kernel configuration. For science cases requiring the maximum \mbox{S/N} at a given spatial resolution, kernel $h$ or larger offers near-ideal coaddition efficiency. For kernel $j$, which delivers the highest spatial resolution, the median gain of $\sim\!77$\% of $\sqrt{9}$ represents the fundamental trade-off between resolution and depth imposed by the LVM fibre size and dither geometry.

%The spatial distribution of $\mathcal{G}_{1\to N}(x,y)$ is shown in Figure~\ref{fig:snrmap} for kernel configurations $j$, $h$, and $f$ and for $N = 3$, 6, and 9 dithers. The gain is highest at the field centre where all dithers overlap and decreases toward the edges. For a given $N$, the gain increases with kernel size as the weight contrast between dithers diminishes. The figure also shows, as a dashed contour, the theoretical $\sqrt{N}$ value in each panel: for kernel $f$ ($\sigma = 140.8''$) the contour covers most of the field, confirming that the large-kernel limit approaches the idealised coadd behaviour. These gain maps are computed directly from the \texttt{Wg} weight arrays stored internally by \textsc{3dcubegen} and can serve as data quality indicators for downstream analyses.

\begin{figure*}[!t]
\centering 
\includegraphics[width=\textwidth]{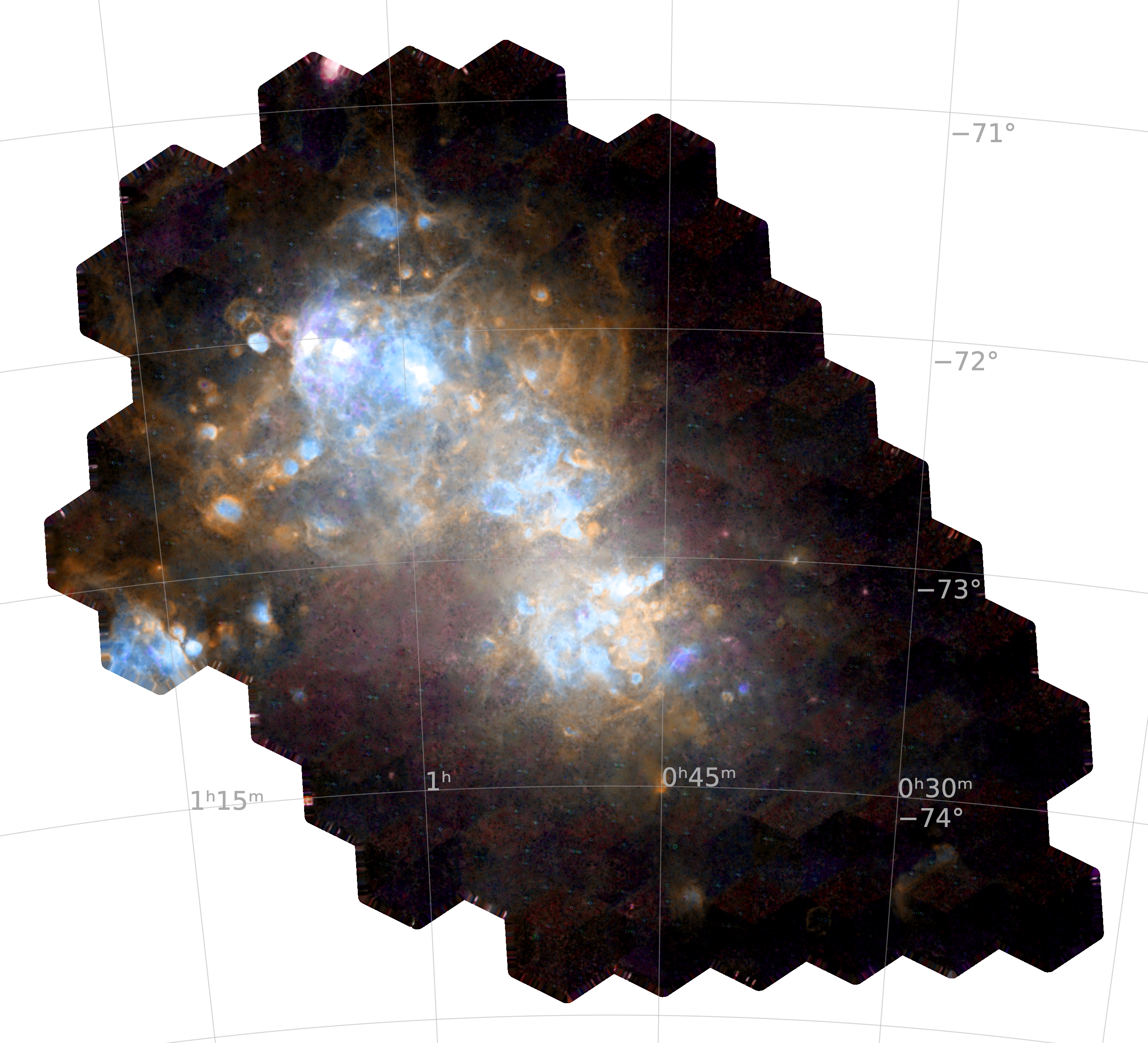}
\caption{Reconstructed colour narrowband image of the full tiled SMC observations using the kernel $j$ configuration ($\sigma=8.8''$, $\alpha=2$). The
RGB composite follows the Hubble palette: \oiii as Blue, \ha as Green, and \sii as red. The H$\alpha$ channel is partially desaturated to enhance the contrast between oxygen- and sulfur-dominated structures, highlighting the complex ionization morphology across the galaxy.}
\label{fig:SMC}
\end{figure*}

\section{The LVM dithered data cubes}

From the analysis in Section \ref{sec:chi}, we can define ten kernel configurations to reconstruct the data cubes of 9-dithered LVM observations using {\sc 3dcubegen}. We show the kernel selection with its minimum spatial resolutions in Table~\ref{tab:resolutions}. In the Table, The spatial resolution is defined as the physical size of the effective PSF dispersion. The kernel configurations $h$, $i$ and $j$ are the only ones whose spatial resolution is limited by the fibre size; for the other configurations, the kernel size limits its spatial resolution. The final products are ten sets of data cubes equal to 1/1.36 kpc to 4.88/6.12 pc resolutions for the LMC/SMC. In addition, using different spatial resolutions not only gives access to explore the physical properties of the observed targets at various scales but can also improve the SNR at lower spatial resolutions due to the number of spectra added per spaxel. 

The data cubes have two extensions, the first in flux in units of $10^{-16} erg\: s^{-2} cm^{-2}$\AA$^{-1}$, and the second extension will contain the propagated error through the entire cube process. For large resolutions, the {\sc 3dcubegen} script can generate the cubes in 7, 13 and 25 wavelength slides across the full LVM spectral range to save memory or create $n\times n$ spatial slides with the entire spectral range for complete spectral analysis. We show the final maps of the LMC (Figure~\ref{fig:LMC}) and SMC (Figure~\ref{fig:SMC}) for the $j$ kernel configuration ($\sigma=8.8''$) that show the actual FoV of the MCs data cubes. For visualization, the reconstructed images are an RGB composition that follows the Hubble palette convention: where \oiii\ emission line is assigned to the blue channel, \ha\ to the green channel, and \sii to the red channel. To enhance the contrast between regions dominated by oxygen and sulphur emission, the green component corresponding to H$\alpha$ was partially desaturated, increasing the visibility of structures traced by [O~III] and [S~II]. This colour treatment highlights the complex ionization structure of the gas and improves the visual interpretation of the different physical conditions across the MCs. While the LVM plans to map the full MCs, we will continuously update the data cubes as new observations are taken. For the data cubes used in this work, we cover the entire centre of the LMC, its northern part and 30 Dorados. In the case of the SMC, we cover its central part and the NGC346 region. Therefore, the actual coverage of the data cubes allows us to explore the stellar population properties from a wide range of resolutions and an integral and resolved point of view.   

\begin{figure*}[!t]
\centering 
\includegraphics[width=1.95\columnwidth]{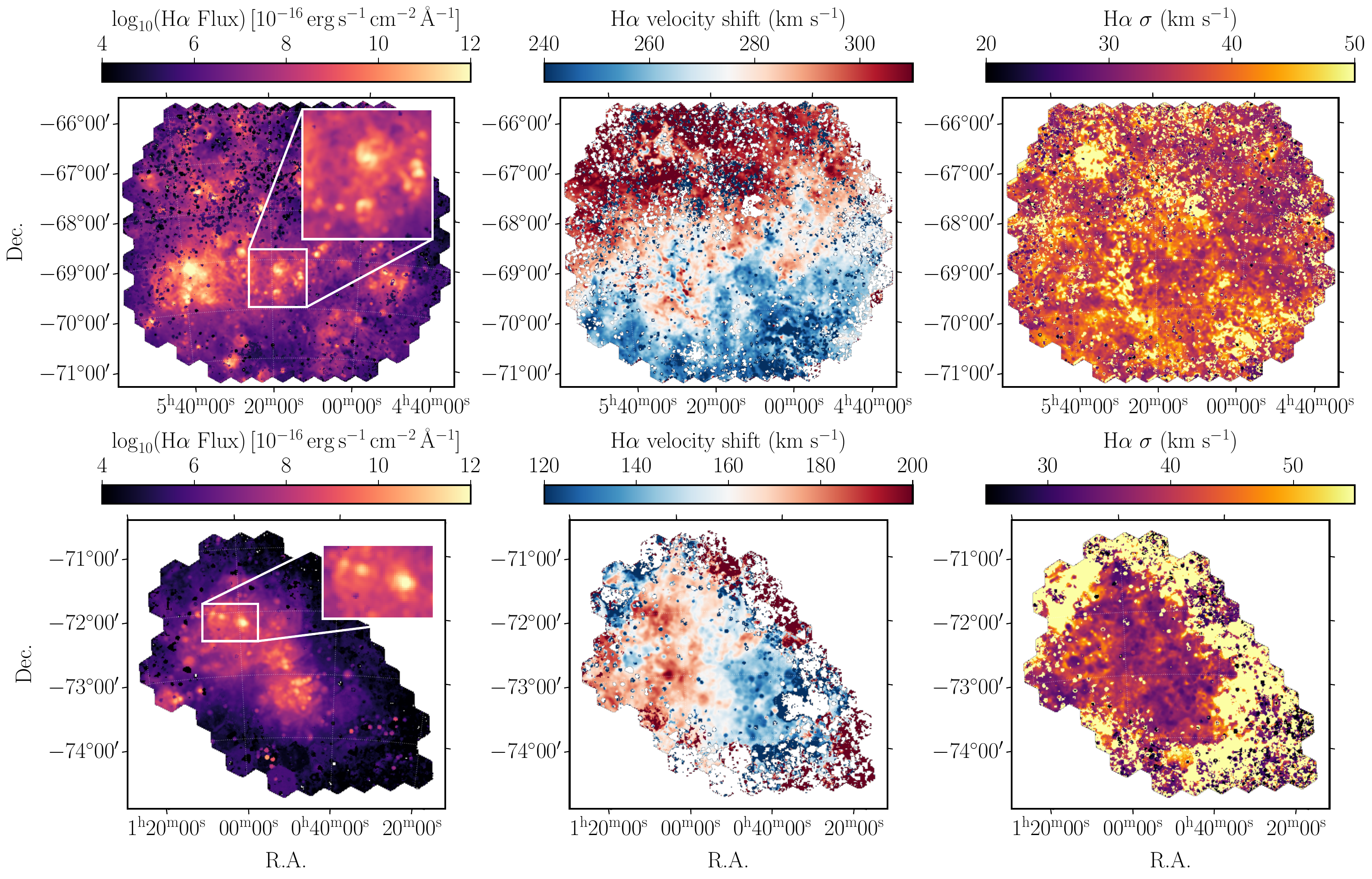}
\caption{H$\alpha$ parametric fits from the {\sc LVM-DAP} using the LMC (upper panels) and SMC (lower panels) data cubes with the $g$ kernel configuration that gives a spatial resolution of 17.05 pc for the LMC and 21.31 pc for the SMC. Left panels shows the total flux maps, the middle panels shows the velocity shift maps, and the right panels shows the velocity dispersions maps.}
\label{fig:DAP}
\end{figure*}

\subsection{Kernel selection for Local Volume targets}
\label{sec:kernel_lv}

While Table~\ref{tab:resolutions} lists the physical resolutions of the ten kernel configurations at the distances of the LMC and SMC, the same configurations apply to any extragalactic LVM target, with the effective physical resolution scaling linearly with distance. For a target at distance $D$, the physical scale corresponding to the effective PSF
dispersion of each configuration is
\begin{equation}
R_{\rm phys}= 2.058
\left(\frac{{\rm FWHM}_{\rm PSF}}{1''}\right)
\left(\frac{D}{1\,{\rm Mpc}}\right)\,{\rm pc},
\label{eq:rphys}
\end{equation}
where ${\rm FWHM}_{\rm PSF}$ is the effective FWHM PSF of each kernel configuration listed in Table~\ref{tab:resolutions}. %This relation allows users to select the optimal configuration for any resolved target in the Local Volume according to the physical scales of interest.

Two regimes can be identified. For targets within the Magellanic system ($D \lesssim 100$~kpc), the configurations $f$, $g$, and $h$ resolve individual \ion{H}{ii} regions and their internal structure at parsec scales. For Local Group and nearby Local Volume galaxies ($D \approx 0.5$--$2$~Mpc), configuration $h$ yields physical resolutions of $\sim$150--600~pc, comparable to the cloud-scale to sub-kpc regimes probed by PHANGS--MUSE \citep[$\sim$50--150~pc;][]{Emsellem2022} and PHANGS--ALMA \citep[$\sim$100~pc;][]{Leroy2021}, enabling direct comparisons of the ionised-gas and stellar-population properties. We emphasise that the oversampled configurations $i$ and $j$ are not advantageous in this regime: as shown in Sections~\ref{sec:chi} and \ref{secc:noise}, these kernels cannot recover spatial structures below the fibre aperture (which at these distances already corresponds to $\sim$85-340~pc) while incurring up to a $\sim$23\% deficit in S/N depth. %Finally, for targets beyond $D \gtrsim 4$~Mpc, the fibre aperture itself corresponds to physical scales larger than $\sim$0.7~kpc, so only global and kpc-scale structures remain accessible, and coarser configurations should be preferred to maximise the coaddition efficiency. 
Therefore, the optimal kernel selection is set by the target distance: the oversampled kernels are only meaningful for the nearest targets, where the fibre-limited resolution translates into parsec scales, while for more distant targets configuration $h$ or larger maximises the spectral depth at no cost in effective spatial resolution.

\section{Implementation with the {\sc LVM-DAP}}\label{sec:dapimp}

The {\sc 3dcubegen} script also generates a new set of RSS from the data cubes containing the 9-dithered spectra coadding as many tiles as needed to be used as an input for the {\sc LVM-DAP} analysis. The advantage of the 9-dither pattern is not only to increase the effective spatial resolution but also to make it possible to coadd the nine exposures to increase the total SNR. Therefore, coadding the spectra through the datacube and then transforming it into a new RSS provides an enhanced way to coadd the data and increase the SNR to get a deeper spectrum before it is analysed by the {\sc LVM-DAP} analysis. The RSS transformation script of {\sc 3dcubegen} also saves the astrometric data and the error spectra. It creates a new fibre map to map spaxels from their original position on the cube with its astrometric position. Therefore, the {\sc LVM-DAP} can use the coadded spectrum to obtain the spatial resolved properties of the stellar populations and the emission line properties of the ionised gas.

In Figure~\ref{fig:DAP} we show an example of the application of the {\sc LVM-DAP} with the full datacubes of the LMC and SMC using the $g$ kernel configuration ($\sigma=2D_{fib}$) of Table \ref{tab:resolutions}. The results of the {\sc LVM-DAP} shows the parametric fit of the kinematics, velocity dispersions and total fluxes with an spatial resolution of 17.05 pc (LMC) and 21.31 pc (SMC), showing all the Star Formation regions, and its diffuse ionized gas (DIG). In a forthcoming set of paper, we will present a full analysis of the nebular emission and its DIG component (Lugo-Aranda et al., in rev), its kinematics (Zerme\~no et al., in prep), its nebular abundances (Casta\~neda et al., in prep), and its resolved/unresolved stellar populations (Zerme\~no et al., in rev and Ibarra-Medel et al., in prep).

In addition, the 2D map reconstruction described in Section~\ref{sec:cubrec}~iv) provides a computationally efficient pathway for kinematic analysis at the best available spatial resolution, particularly relevant for science cases where the spatial coherence of individual fibre measurements is more important than the improved \mbox{S/N} of the coadded cube. One such case is the mapping of the difference between gas and stellar velocity fields across the LMC and SMC, which is expected to carry signatures of their past interaction \citep{Besla+2012, Choi+2022}: the collisional gas would have been more strongly perturbed than the collisionless stars during the \textsc{smc}--\textsc{lmc} encounter, producing a spatially coherent offset between $V_\mathrm{gas}$ and $V_\mathrm{stars}$ that the \textsc{lvm} data are perfectly positioned to map at parsec-scale resolution. This analysis will be presented in forthcoming work (Zerme\~{n}o et al., in prep).

\section{Summary and Conclusions}

%In this work we describe the methodology for the datacube reconstruction for the dithered data of the SDSS-V LVM.  We perform an analysis to quantify how well it is possible to recover the most spatial information depending on the selection of the kernel of the 2D interpolation by measuring the differences between the reconstructed image and the DSS images. We also implement a 2D deconvolution to measure the effective FWHM-PSF on the reconstructed image using our methodology. We describe our main conclusions as follows:

In this work, we describe a methodology for data cube reconstruction of the dithered data from the SDSS-V LVM. We performed an analysis to quantify how well our algorithm recovers spatial information as a function of the kernel selection used in the 2D interpolation. This was done by measuring the differences between the reconstructed images and an archival DSS image of the galaxy NGC 1365. This approach allowed us to characterise the noise properties of the reconstructed cubes. We also implement a 2D deconvolution to measure the effective FWHM-PSF on the reconstructed image using our methodology and quantify the \mbox{S/N} improvement achieved by the 9-point dither strategy. Our main conclusions are as follows:

\begin{itemize}
    \item[a)] We found that there is a minimum kernel size to avoid oversampling during the 2D interpolation. This minimum size can be fixed at $\sigma=1/4D_{fib}$ with $\alpha=2$ and any other combination of the kernel that retrieves an equivalent kernel shape. When the 2D interpolation is oversampled, the minimum spatial resolution is determined by the fibre size.
    \item[b)] We found that for Kernel values between $\sigma=2D_{fib}$ and $\sigma=1/4D_{fib}$ show that the effective FWHM-PSF of the reconstructed image approaches the relation $\sqrt{D_{fig}^2 +8\ln2\sigma^2}$. This implies that the  spatial resolution of the reconstructed image is limited by combined contribution of the fibre size and the kernel size, even in the absence of oversampling. Therefore, the minimum spatial resolution is ultimately limited by the fibre size, regardless of the kernel size adopted.   
    \item[c)] Deconvolution of the reconstructed images using an empirical Gaussian kernel generates spurious structures and does not reliably recover spatial information below the effective fibre aperture. This conclusion applies specifically to the Gaussian approximation adopted here. Improved results may be achieved through deconvolution using a physically motivated kernel constructed from the convolution of the 35.3$''$ circular fibre aperture and the interpolation kernel $w(r)$ \citep{vanderMarel1995,vanderMarel1997}.
    %\item[c)] The deconvolution of the reconstructed images generates spurious structures, and therefore this work don't recommend to implement on the reconstructed datacubes. 
    %\item[d)] For one-dither data, we recommend to use a kernel size larger or equal of two times the fibre diameter  $\sigma\geq D_{fib}$ to deal with the missing data due to the IFU fill factor.

    \item[d)] The per-spaxel error extension of each data cube correctly propagates the formal inverse-variance uncertainties from the \textsc{lvm-drp} RSS spectra. However, the kernel interpolation introduces spatial covariance between neighbouring spaxels over a scale $r \lesssim 2\sigma$, which is not encoded in the stored per-spaxel errors. This covariance is characterised by the 2-point correlation function $C(\Delta r)$, which follows a self-similar profile that scales with $\sigma$ across all kernel configurations. Users performing aperture or bin-level spectral extractions should apply an empirical noise rescaling to account for this effect; the full covariance tensor $\boldsymbol{\varepsilon}^{N,M,N,M}$ can optionally be saved via the \texttt{fcovmat=True} option.
    
    \item[e)] The \mbox{S/N} improvement from combining $N$ dithers is not a fixed $\sqrt{N}$ factor but a kernel-dependent, spatially varying quantity $\mathcal{G}_{1\to N}(x,y) = (S_N/S_1)\sqrt{Q_1/Q_N}$. For kernels $\sigma \gtrsim D_\mathrm{fib}$ (configurations $h$ and $f$), the median gain reaches $\sim\!100$\% of $\sqrt{N}$ across the field, confirming near-ideal coaddition efficiency. For kernel $j$ ($\sigma = 8.8''$), the median gain is $\sim\!77$\% of $\sqrt{N}$, reflecting the fundamental trade-off between spatial resolution and \mbox{S/N} depth imposed by the LVM fibre size and dither geometry.

    \item[f)] We define a set of 10 kernel configurations spanning spatial resolutions from $\sim\!1.1$ kpc to $\sim\!4.9$ pc at the distance of the LMC, enabling multi-scale analysis of the ionised gas, stellar populations, and kinematics of the MC and LV. The physical resolution of each configuration scales linearly with target distance (Equation~\ref{eq:rphys}); for targets beyond the Magellanic system, the oversampled configurations provide no additional spatial information while reducing the S/N depth, and we therefore recommend reconstructing at the fibre-sampling limit (configuration $h$) or larger. %The resulting data cubes will be publicly available and continuously updated as new \textsc{lvm} observations are acquired.
    %\item[d)] We define a set of 10 kernel configurations to reconstruct the dithered LVM data into a set of datacubes. By this way, it is possible to use all the 9-dithered exposures to generate a deeper spectra in comparison with the single exposure spectra to be analysed by the LVM-DAP.
\end{itemize}

\section{ACKNOWLEDGEMENTS}
This work was supported by UNAM PASPA – DGAPA. H.I.M. acknowledges the support from grant CONAHCyT CBF2023-2024-1418, IN-106823 PAPIIT UNAM, IN-119123 PAPIT UNAM and CONHACyT CF-2023-G-543. AZL-A gratefully acknowledges the support provided by the Postdoctoral Program (POSDOC) of UNAM (Universidad Nacional Autónoma de México). R.Z. would like to express his appreciation for the support provided by SECIHTI (Secretaría de Ciencia, Humanidades, Tecnología e Innovación) in form of their Posdoctoral Grant. A.W. gratefully acknowledges the support provided by PAPIIT IN107725. C.R-Z acknowledges support by project IN107226 PAPIIT UNAM. G.A.B. acknowledges the support from the ANID Basal project FB210003. 

Funding for the Sloan Digital Sky Survey V has been provided by the Alfred P. Sloan Foundation, the Heising-Simons Foundation, the National Science Foundation, and the Participating Institutions. SDSS acknowledges support and resources from the Center for High-Performance Computing at the University of Utah. SDSS telescopes are located at Apache Point Observatory, funded by the Astrophysical Research Consortium and operated by New Mexico State University, and at Las Campanas Observatory, operated by the Carnegie Institution for Science. The SDSS web site is \url{www.sdss.org}.

SDSS is managed by the Astrophysical Research Consortium for the Participating Institutions of the SDSS Collaboration, including the Carnegie Institution for Science, Chilean National Time Allocation Committee (CNTAC) ratified researchers, Caltech, the Gotham Participation Group, Harvard University, Heidelberg University, The Flatiron Institute, The Johns Hopkins University, L'Ecole polytechnique f\'{e}d\'{e}rale de Lausanne (EPFL), Leibniz-Institut f\"{u}r Astrophysik Potsdam (AIP), Max-Planck-Institut f\"{u}r Astronomie (MPIA Heidelberg), Max-Planck-Institut f\"{u}r Extraterrestrische Physik (MPE), Nanjing University, National Astronomical Observatories of China (NAOC), New Mexico State University, The Ohio State University, Pennsylvania State University, Smithsonian Astrophysical Observatory, Space Telescope Science Institute (STScI), the Stellar Astrophysics Participation Group, Universidad Nacional Aut\'{o}noma de M\'{e}xico, University of Arizona, University of Colorado Boulder, University of Illinois at Urbana-Champaign, University of Toronto, University of Utah, University of Virginia, Yale University, and Yunnan University.

%----------------------------------------------------------
% Al final del documento, antes de \end{document}
%----------------------------------------------------------
%\section{APPENDICES}

%\appendix

%\section{One-dither reconstruction}

%\begin{figure*}
%\includegraphics[width=2\columnwidth]{figures/optionF.pdf}
%\includegraphics[width=2\columnwidth]{figures/optionG.pdf}
%\includegraphics[width=2\columnwidth]{figures/optionH.pdf}
%\caption{Reconstructed images from the 2D interpolation process of NGC 1365 using one-dither observation. From top to bottom panels: case for $\sigma=8.8''$ ($1/4D_{fib}$, upper panel), case for $\sigma=17.6.8''$ ($1/2D_{fib}$, middle panel), case for $\sigma=35.2.6''$ ($D_{fib}$, bottom panel). The value of $\alpha=2$ is constant for all cases. From left to right panels: 2D reconstructed image in the R photometric band using the one-dithered exposures of the LVM data (left), DSS2 image in the R photometric band of the same region (middle), and the residual image from both images (left). The colour scale and bar are in log $erg/s/cm^2/$\AA\ for all panels.}
%\label{fig:one-recons}
%\end{figure*}

%\begin{figure*}
%\includegraphics[width=1\columnwidth]{figures/deconvchiDb.pdf}
%\includegraphics[width=1\columnwidth]{figures/deconvchiCb.pdf}
%\includegraphics[width=1\columnwidth]{figures/deconvchiBb.pdf}
%\includegraphics[width=1\columnwidth]{figures/deconvchiAb.pdf}
%\caption{}
%\label{fig:one-recons2}
%\end{figure*}

%\bibliographystyle{rasti}
\renewcommand{\refname}{REFERENCES}
\bibliography{example}

\begin{thebibliography}{}
\expandafter\ifx\csname natexlab\endcsname\relax\def\natexlab#1{#1}\fi
\providecommand{\url}[1]{\href{#1}{#1}}
\providecommand{\dodoi}[1]{doi:~\href{http://doi.org/#1}{\nolinkurl{#1}}}
\providecommand{\doeprint}[1]{\href{http://ascl.net/#1}{\nolinkurl{http://ascl.net/#1}}}
\providecommand{\doarXiv}[1]{\href{https://arxiv.org/abs/#1}{\nolinkurl{https://arxiv.org/abs/#1}}}

\bibitem[{{Abdurro'uf} {et~al.}(2022){Abdurro'uf}, {Accetta}, {Aerts}, {Silva
  Aguirre}, {Ahumada}, {Ajgaonkar}, {Filiz Ak}, {Alam}, {Allende Prieto},
  {Almeida}, {Anders}, {Anderson}, {Andrews}, {Anguiano}, {Aquino-Ort{\'\i}z},
  {Arag{\'o}n-Salamanca}, {Argudo-Fern{\'a}ndez}, {Ata}, {Aubert},
  {Avila-Reese}, {Badenes}, {Barb{\'a}}, {Barger}, {Barrera-Ballesteros},
  {Beaton}, {Beers}, {Belfiore}, {Bender}, {Bernardi}, {Bershady}, {Beutler},
  {Bidin}, {Bird}, {Bizyaev}, {Blanc}, {Blanton}, {Boardman}, {Bolton},
  {Boquien}, {Borissova}, {Bovy}, {Brandt}, {Brown}, {Brownstein}, {Brusa},
  {Buchner}, {Bundy}, {Burchett}, {Bureau}, {Burgasser}, {Cabang}, {Campbell},
  {Cappellari}, {Carlberg}, {Wanderley}, {Carrera}, {Cash}, {Chen}, {Chen},
  {Cherinka}, {Chiappini}, {Choi}, {Chojnowski}, {Chung}, {Clerc}, {Cohen},
  {Comerford}, {Comparat}, {da Costa}, {Covey}, {Crane}, {Cruz-Gonzalez},
  {Culhane}, {Cunha}, {Dai}, {Damke}, {Darling}, {Davidson}, {Davies},
  {Dawson}, {De Lee}, {Diamond-Stanic}, {Cano-D{\'\i}az}, {S{\'a}nchez},
  {Donor}, {Duckworth}, {Dwelly}, {Eisenstein}, {Elsworth}, {Emsellem},
  {Eracleous}, {Escoffier}, {Fan}, {Farr}, {Feng}, {Fern{\'a}ndez-Trincado},
  {Feuillet}, {Filipp}, {Fillingham}, {Frinchaboy}, {Fromenteau}, {Galbany},
  {Garc{\'\i}a}, {Garc{\'\i}a-Hern{\'a}ndez}, {Ge}, {Geisler}, {Gelfand},
  {G{\'e}ron}, {Gibson}, {Goddy}, {Godoy-Rivera}, {Grabowski}, {Green},
  {Greener}, {Grier}, {Griffith}, {Guo}, {Guy}, {Hadjara}, {Harding},
  {Hasselquist}, {Hayes}, {Hearty}, {Hern{\'a}ndez}, {Hill}, {Hogg},
  {Holtzman}, {Horta}, {Hsieh}, {Hsu}, {Hsu}, {Huber}, {Huertas-Company},
  {Hutchinson}, {Hwang}, {Ibarra-Medel}, {Chitham}, {Ilha}, {Imig}, {Jaekle},
  {Jayasinghe}, {Ji}, {Johnson}, {Jones}, {J{\"o}nsson}, {Katkov}, {Khalatyan},
  {Kinemuchi}, {Kisku}, {Knapen}, {Kneib}, {Kollmeier}, {Kong}, {Kounkel},
  {Kreckel}, {Krishnarao}, {Lacerna}, {Lane}, {Langgin}, {Lavender}, {Law},
  {Lazarz}, {Leung}, {Leung}, {Lewis}, {Li}, {Li}, {Lian}, {Liang}, {Lin},
  {Lin}, {Lin}, {Lintott}, {Long}, {Longa-Pe{\~n}a}, {L{\'o}pez-Cob{\'a}},
  {Lu}, {Lundgren}, {Luo}, {Mackereth}, {de la Macorra}, {Mahadevan},
  {Majewski}, {Manchado}, {Mandeville}, {Maraston}, {Margalef-Bentabol},
  {Masseron}, {Masters}, {Mathur}, {McDermid}, {Mckay}, {Merloni},
  {Merrifield}, {Meszaros}, {Miglio}, {Di Mille}, {Minniti}, {Minsley},
  {Monachesi}, {Moon}, {Mosser}, {Mulchaey}, {Muna}, {Mu{\~n}oz}, {Myers},
  {Myers}, {Nadathur}, {Nair}, {Nandra}, {Neumann}, {Newman}, {Nidever},
  {Nikakhtar}, {Nitschelm}, {O'Connell}, {Garma-Oehmichen}, {Luan Souza de
  Oliveira}, {Olney}, {Oravetz}, {Ortigoza-Urdaneta}, {Osorio}, {Otter},
  {Pace}, {Padilla}, {Pan}, {Pan}, {Parikh}, {Parker}, {Peirani}, {Pe{\~n}a
  Ram{\'\i}rez}, {Penny}, {Percival}, {Perez-Fournon}, {Pinsonneault},
  {Poidevin}, {Poovelil}, {Price-Whelan}, {B{\'a}rbara de Andrade Queiroz},
  {Raddick}, {Ray}, {Rembold}, {Riddle}, {Riffel}, {Riffel}, {Rix}, {Robin},
  {Rodr{\'\i}guez-Puebla}, {Roman-Lopes}, {Rom{\'a}n-Z{\'u}{\~n}iga}, {Rose},
  {Ross}, {Rossi}, {Rubin}, {Salvato}, {S{\'a}nchez}, {S{\'a}nchez-Gallego},
  {Sanderson}, {Santana Rojas}, {Sarceno}, {Sarmiento}, {Sayres}, {Sazonova},
  {Schaefer}, {Schiavon}, {Schlegel}, {Schneider}, {Schultheis}, {Schwope},
  {Serenelli}, {Serna}, {Shao}, {Shapiro}, {Sharma}, {Shen}, {Shetrone}, {Shu},
  {Simon}, {Skrutskie}, {Smethurst}, {Smith}, {Sobeck}, {Spoo}, {Sprague},
  {Stark}, {Stassun}, {Steinmetz}, {Stello}, {Stone-Martinez},
  {Storchi-Bergmann}, {Stringfellow}, {Stutz}, {Su}, {Taghizadeh-Popp},
  {Talbot}, {Tayar}, {Telles}, {Teske}, {Thakar}, {Theissen}, {Tkachenko},
  {Thomas}, {Tojeiro}, {Hernandez Toledo}, {Troup}, {Trump}, {Trussler},
  {Turner}, {Tuttle}, {Unda-Sanzana}, {V{\'a}zquez-Mata}, {Valentini},
  {Valenzuela}, {Vargas-Gonz{\'a}lez}, {Vargas-Maga{\~n}a}, {Alfaro},
  {Villanova}, {Vincenzo}, {Wake}, {Warfield}, {Washington}, {Weaver},
  {Weijmans}, {Weinberg}, {Weiss}, {Westfall}, {Wild}, {Wilde}, {Wilson},
  {Wilson}, {Wilson}, {Wolf}, {Wood-Vasey}, {Yan}, {Zamora}, {Zasowski},
  {Zhang}, {Zhao}, {Zheng}, {Zheng}, \& {Zhu}}]{Abdurro+22}
{Abdurro'uf}, {Accetta}, K., {Aerts}, C., {et~al.} 2022, \apjs, 259, 35,
  \dodoi{10.3847/1538-4365/ac4414}

\bibitem[{{Almeida} {et~al.}(2023){Almeida}, {Anderson},
  {Argudo-Fern{\'a}ndez}, {Badenes}, {Barger}, {Barrera-Ballesteros}, {Bender},
  {Benitez}, {Besser}, {Bird}, {Bizyaev}, {Blanton}, {Bochanski}, {Bovy},
  {Brandt}, {Brownstein}, {Buchner}, {Bulbul}, {Burchett}, {Cano D{\'\i}az},
  {Carlberg}, {Casey}, {Chandra}, {Cherinka}, {Chiappini}, {Coker}, {Comparat},
  {Conroy}, {Contardo}, {Cortes}, {Covey}, {Crane}, {Cunha}, {Dabbieri},
  {Davidson}, {Davis}, {de Andrade Queiroz}, {De Lee}, {M{\'e}ndez Delgado},
  {Demasi}, {Di Mille}, {Donor}, {Dow}, {Dwelly}, {Eracleous}, {Eriksen},
  {Fan}, {Farr}, {Frederick}, {Fries}, {Frinchaboy}, {G{\"a}nsicke}, {Ge},
  {Gonz{\'a}lez {\'A}vila}, {Grabowski}, {Grier}, {Guiglion}, {Gupta}, {Hall},
  {Hawkins}, {Hayes}, {Hermes}, {Hern{\'a}ndez-Garc{\'\i}a}, {Hogg},
  {Holtzman}, {Ibarra-Medel}, {Ji}, {Jofre}, {Johnson}, {Jones}, {Kinemuchi},
  {Kluge}, {Koekemoer}, {Kollmeier}, {Kounkel}, {Krishnarao}, {Krumpe},
  {Lacerna}, {Lago}, {Laporte}, {Liu}, {Liu}, {Liu}, {Lopes}, {Macktoobian},
  {Majewski}, {Malanushenko}, {Maoz}, {Masseron}, {Masters}, {Matijevic},
  {McBride}, {Medan}, {Merloni}, {Morrison}, {Myers}, {M{\'e}sz{\'a}ros},
  {Negrete}, {Nidever}, {Nitschelm}, {Oravetz}, {Oravetz}, {Pan}, {Peng},
  {Pinsonneault}, {Pogge}, {Qiu}, {Ramirez}, {Rix}, {Fern{\'a}ndez Rosso},
  {Runnoe}, {Salvato}, {Sanchez}, {Santana}, {Saydjari}, {Sayres},
  {Schlaufman}, {Schneider}, {Schwope}, {Serna}, {Shen}, {Sobeck}, {Song},
  {Souto}, {Spoo}, {Stassun}, {Steinmetz}, {Straumit}, {Stringfellow},
  {S{\'a}nchez-Gallego}, {Taghizadeh-Popp}, {Tayar}, {Thakar}, {Tissera},
  {Tkachenko}, {Hernandez Toledo}, {Trakhtenbrot}, {Fern{\'a}ndez-Trincado},
  {Troup}, {Trump}, {Tuttle}, {Ulloa}, {Vazquez-Mata}, {Vera Alfaro},
  {Villanova}, {Wachter}, {Weijmans}, {Wheeler}, {Wilson}, {Wojno}, {Wolf},
  {Xue}, {Ybarra}, {Zari}, \& {Zasowski}}]{Almeida+2023}
{Almeida}, A., {Anderson}, S.~F., {Argudo-Fern{\'a}ndez}, M., {et~al.} 2023,
  \apjs, 267, 44, \dodoi{10.3847/1538-4365/acda98}

\bibitem[{{Aquino-Ort{\'\i}z} {et~al.}(2020){Aquino-Ort{\'\i}z}, {S{\'a}nchez},
  {Valenzuela}, {Hern{\'a}ndez-Toledo}, {Jin}, {Zhu}, {van de Ven},
  {Barrera-Ballesteros}, {Avila-Reese}, {Rodr{\'\i}guez-Puebla}, \&
  {Tissera}}]{Aquino+2020}
{Aquino-Ort{\'\i}z}, E., {S{\'a}nchez}, S.~F., {Valenzuela}, O., {et~al.} 2020,
  \apj, 900, 109, \dodoi{10.3847/1538-4357/aba94e}

\bibitem[{{Avila-Reese} {et~al.}(2023){Avila-Reese}, {Ibarra-Medel}, {Lacerna},
  {Rodr{\'\i}guez-Puebla}, {V{\'a}zquez-Mata}, {S{\'a}nchez},
  {Hern{\'a}ndez-Toledo}, \& {Cannarozzo}}]{Avila-Reese+2023}
{Avila-Reese}, V., {Ibarra-Medel}, H., {Lacerna}, I., {et~al.} 2023, \mnras,
  523, 4251, \dodoi{10.1093/mnras/stad1638}

\bibitem[{{Bacon} {et~al.}(2010){Bacon}, {Accardo}, {Adjali}, {Anwand},
  {Bauer}, {Biswas}, {Blaizot}, {Boudon}, {Brau-Nogue}, {Brinchmann},
  {Caillier}, {Capoani}, {Carollo}, {Contini}, {Couderc}, {Daguis{\'e}},
  {Deiries}, {Delabre}, {Dreizler}, {Dubois}, {Dupieux}, {Dupuy}, {Emsellem},
  {Fechner}, {Fleischmann}, {Fran{\c{c}}ois}, {Gallou}, {Gharsa}, {Glindemann},
  {Gojak}, {Guiderdoni}, {Hansali}, {Hahn}, {Jarno}, {Kelz}, {Koehler},
  {Kosmalski}, {Laurent}, {Le Floch}, {Lilly}, {Lizon}, {Loupias}, {Manescau},
  {Monstein}, {Nicklas}, {Olaya}, {Pares}, {Pasquini}, {P{\'e}contal-Rousset},
  {Pell{\'o}}, {Petit}, {Popow}, {Reiss}, {Remillieux}, {Renault}, {Roth},
  {Rupprecht}, {Serre}, {Schaye}, {Soucail}, {Steinmetz}, {Streicher}, {Stuik},
  {Valentin}, {Vernet}, {Weilbacher}, {Wisotzki}, \& {Yerle}}]{Bacon+2010}
{Bacon}, R., {Accardo}, M., {Adjali}, L., {et~al.} 2010, in Society of
  Photo-Optical Instrumentation Engineers (SPIE) Conference Series, Vol. 7735,
  Ground-based and Airborne Instrumentation for Astronomy III, ed. I.~S.
  {McLean}, S.~K. {Ramsay}, \& H.~{Takami}, 773508, \dodoi{10.1117/12.856027}

\bibitem[{{Ben{\'\i}tez} {et~al.}(2023){Ben{\'\i}tez}, {Ibarra-Medel},
  {Negrete}, {Cruz-Gonz{\'a}lez}, {Rodr{\'\i}guez-Espinosa}, {Liu}, \&
  {Shen}}]{Benitez+2023}
{Ben{\'\i}tez}, E., {Ibarra-Medel}, H., {Negrete}, C.~A., {et~al.} 2023, \apj,
  952, 45, \dodoi{10.3847/1538-4357/acce3e}

\bibitem[{{Besla} {et~al.}(2012){Besla}, {Kallivayalil}, {Hernquist}, {van der
  Marel}, {Cox}, \& {Kere{\v{s}}}}]{Besla+2012}
{Besla}, G., {Kallivayalil}, N., {Hernquist}, L., {et~al.} 2012, \mnras, 421,
  2109, \dodoi{10.1111/j.1365-2966.2012.20466.x}

\bibitem[{{Bundy} {et~al.}(2015){Bundy}, {Bershady}, {Law}, {Yan}, {Drory},
  {MacDonald}, {Wake}, {Cherinka}, {S{\'a}nchez-Gallego}, {Weijmans}, {Thomas},
  {Tremonti}, {Masters}, {Coccato}, {Diamond-Stanic}, {Arag{\'o}n-Salamanca},
  {Avila-Reese}, {Badenes}, {Falc{\'o}n-Barroso}, {Belfiore}, {Bizyaev},
  {Blanc}, {Bland-Hawthorn}, {Blanton}, {Brownstein}, {Byler}, {Cappellari},
  {Conroy}, {Dutton}, {Emsellem}, {Etherington}, {Frinchaboy}, {Fu}, {Gunn},
  {Harding}, {Johnston}, {Kauffmann}, {Kinemuchi}, {Klaene}, {Knapen},
  {Leauthaud}, {Li}, {Lin}, {Maiolino}, {Malanushenko}, {Malanushenko}, {Mao},
  {Maraston}, {McDermid}, {Merrifield}, {Nichol}, {Oravetz}, {Pan}, {Parejko},
  {Sanchez}, {Schlegel}, {Simmons}, {Steele}, {Steinmetz}, {Thanjavur},
  {Thompson}, {Tinker}, {van den Bosch}, {Westfall}, {Wilkinson}, {Wright},
  {Xiao}, \& {Zhang}}]{Bundy+2015}
{Bundy}, K., {Bershady}, M.~A., {Law}, D.~R., {et~al.} 2015, \apj, 798, 7,
  \dodoi{10.1088/0004-637X/798/1/7}

\bibitem[{{Camps-Fari{\~n}a} {et~al.}(2022){Camps-Fari{\~n}a}, {S{\'a}nchez},
  {Mej{\'\i}a-Narv{\'a}ez}, {Lacerda}, {Carigi}, {Bruzual}, {Alvarez-Hurtado},
  {Drory}, {Lane}, {Boardman}, \& {Blanc}}]{Artemi+2022}
{Camps-Fari{\~n}a}, A., {S{\'a}nchez}, S.~F., {Mej{\'\i}a-Narv{\'a}ez}, A.,
  {et~al.} 2022, \apj, 933, 44, \dodoi{10.3847/1538-4357/ac6cea}

\bibitem[{{Cano-D{\'\i}az} {et~al.}(2022){Cano-D{\'\i}az},
  {Hern{\'a}ndez-Toledo}, {Rodr{\'\i}guez-Puebla}, {Ibarra-Medel},
  {{\'A}vila-Reese}, {Valenzuela}, {Medellin-Hurtado}, {V{\'a}zquez-Mata},
  {Weijmans}, {Gonz{\'a}lez}, {Aquino-Ortiz}, {Mart{\'\i}nez-V{\'a}zquez}, \&
  {Lane}}]{Cano-Diaz+22}
{Cano-D{\'\i}az}, M., {Hern{\'a}ndez-Toledo}, H.~M., {Rodr{\'\i}guez-Puebla},
  A., {et~al.} 2022, \aj, 164, 127, \dodoi{10.3847/1538-3881/ac8549}

\bibitem[{{Cappellari} \& {Copin}(2003)}]{Cappellari+2003}
{Cappellari}, M., \& {Copin}, Y. 2003, \mnras, 342, 345,
  \dodoi{10.1046/j.1365-8711.2003.06541.x}

\bibitem[{{Cappellari} {et~al.}(2011){Cappellari}, {Emsellem}, {Krajnovi{\'c}},
  {McDermid}, {Scott}, {Verdoes Kleijn}, {Young}, {Alatalo}, {Bacon}, {Blitz},
  {Bois}, {Bournaud}, {Bureau}, {Davies}, {Davis}, {de Zeeuw}, {Duc},
  {Khochfar}, {Kuntschner}, {Lablanche}, {Morganti}, {Naab}, {Oosterloo},
  {Sarzi}, {Serra}, \& {Weijmans}}]{Cappellari+2011}
{Cappellari}, M., {Emsellem}, E., {Krajnovi{\'c}}, D., {et~al.} 2011, \mnras,
  413, 813, \dodoi{10.1111/j.1365-2966.2010.18174.x}

\bibitem[{{Choi} {et~al.}(2022){Choi}, {Olsen}, {Besla}, {van der Marel},
  {Zivick}, {Kallivayalil}, \& {Nidever}}]{Choi+2022}
{Choi}, Y., {Olsen}, K. A.~G., {Besla}, G., {et~al.} 2022, \apj, 927, 153,
  \dodoi{10.3847/1538-4357/ac4e90}

\bibitem[{{Croom} {et~al.}(2012){Croom}, {Lawrence}, {Bland-Hawthorn},
  {Bryant}, {Fogarty}, {Richards}, {Goodwin}, {Farrell}, {Miziarski}, {Heald},
  {Jones}, {Lee}, {Colless}, {Brough}, {Hopkins}, {Bauer}, {Birchall}, {Ellis},
  {Horton}, {Leon-Saval}, {Lewis}, {L{\'o}pez-S{\'a}nchez}, {Min}, {Trinh}, \&
  {Trowland}}]{Croom+2012}
{Croom}, S.~M., {Lawrence}, J.~S., {Bland-Hawthorn}, J., {et~al.} 2012, \mnras,
  421, 872, \dodoi{10.1111/j.1365-2966.2011.20365.x}

\bibitem[{{Drory} {et~al.}(2024){Drory}, {Blanc}, {Kreckel}, {S{\'a}nchez},
  {Mej{\'\i}a-Narv{\'a}ez}, {Johnston}, {Jones}, {Pellegrini}, {Konidaris},
  {Herbst}, {S{\'a}nchez-Gallego}, {Kollmeier}, {de Almeida},
  {Barrera-Ballesteros}, {Bizyaev}, {Brownstein}, {i Saguer}, {Cherinka},
  {Cioni}, {Congiu}, {Cosens}, {Dias}, {Donor}, {Egorov}, {Egorova}, {Froning},
  {Garc{\'\i}a}, {Glover}, {Greve}, {H{\"a}berle}, {Hoy}, {Ibarra}, {Li},
  {Klessen}, {Krishnarao}, {Kumari}, {Long}, {M{\'e}ndez-Delgado}, {Popa},
  {Ramirez}, {Rix}, {S{\'a}nchez}, {Sankrit}, {Sattler}, {Sayres}, {Singh},
  {Stringfellow}, {Wachter}, {Watkins}, {Wong}, \& {Wofford}}]{Drory+2024}
{Drory}, N., {Blanc}, G.~A., {Kreckel}, K., {et~al.} 2024, \aj, 168, 198,
  \dodoi{10.3847/1538-3881/ad6de9}

\bibitem[{{Emsellem} {et~al.}(2022){Emsellem}, {Schinnerer}, {Santoro},
  {Belfiore}, {Pessa}, {McElroy}, {Blanc}, {Congiu}, {Groves}, {Ho}, {Kreckel},
  {Razza}, {Sanchez-Blazquez}, {Egorov}, {Faesi}, {Klessen}, {Leroy}, {Meidt},
  {Querejeta}, {Rosolowsky}, {Scheuermann}, {Anand}, {Barnes},
  {Be{\v{s}}li{\'c}}, {Bigiel}, {Boquien}, {Cao}, {Chevance}, {Dale},
  {Eibensteiner}, {Glover}, {Grasha}, {Henshaw}, {Hughes}, {Koch}, {Kruijssen},
  {Lee}, {Liu}, {Pan}, {Pety}, {Saito}, {Sandstrom}, {Schruba}, {Sun},
  {Thilker}, {Usero}, {Watkins}, \& {Williams}}]{Emsellem2022}
{Emsellem}, E., {Schinnerer}, E., {Santoro}, F., {et~al.} 2022, \aap, 659,
  A191, \dodoi{10.1051/0004-6361/202141727}

\bibitem[{{Feger} {et~al.}(2020){Feger}, {Case}, {Zhelem}, {Kripak},
  {Lawrence}, {Schwab}, {Herbst}, {Blanc}, {Bilgi}, {Konidaris}, {Hebert},
  {Wachter}, {Ramirez}, {Drory}, \& {Froning}}]{Feger+2020}
{Feger}, T., {Case}, S., {Zhelem}, R., {et~al.} 2020, in Society of
  Photo-Optical Instrumentation Engineers (SPIE) Conference Series, Vol. 11447,
  Ground-based and Airborne Instrumentation for Astronomy VIII, ed. C.~J.
  {Evans}, J.~J. {Bryant}, \& K.~{Motohara}, 114478N,
  \dodoi{10.1117/12.2562185}

\bibitem[{{Gonz{\'a}lez Delgado} {et~al.}(2017){Gonz{\'a}lez Delgado},
  {P{\'e}rez}, {Cid Fernandes}, {Garc{\'\i}a-Benito}, {L{\'o}pez
  Fern{\'a}ndez}, {Vale Asari}, {Cortijo-Ferrero}, {de Amorim}, {Lacerda},
  {S{\'a}nchez}, {Lehnert}, \& {Walcher}}]{Gonzalez-Delgado+2017}
{Gonz{\'a}lez Delgado}, R.~M., {P{\'e}rez}, E., {Cid Fernandes}, R., {et~al.}
  2017, \aap, 607, A128, \dodoi{10.1051/0004-6361/201730883}

\bibitem[{{Grandmont} {et~al.}(2012){Grandmont}, {Drissen}, {Mandar},
  {Thibault}, \& {Baril}}]{Grandmont+2012}
{Grandmont}, F., {Drissen}, L., {Mandar}, J., {Thibault}, S., \& {Baril}, M.
  2012, in Society of Photo-Optical Instrumentation Engineers (SPIE) Conference
  Series, Vol. 8446, Ground-based and Airborne Instrumentation for Astronomy
  IV, ed. I.~S. {McLean}, S.~K. {Ramsay}, \& H.~{Takami}, 84460U,
  \dodoi{10.1117/12.926782}

\bibitem[{{Herbst} {et~al.}(2020){Herbst}, {Bilgi}, {Bizenberger}, {Blanc},
  {Briegel}, {Case}, {Drory}, {Feger}, {Froning}, {Gaessler}, {Hebert},
  {Konidaris}, {Lanz}, {Mohr}, {Pak}, {Ram{\'\i}rez}, {Rohloff},
  {S{\'a}nchez-Gallego}, \& {Wachter}}]{Herbst+2020}
{Herbst}, T.~M., {Bilgi}, P., {Bizenberger}, P., {et~al.} 2020, in Society of
  Photo-Optical Instrumentation Engineers (SPIE) Conference Series, Vol. 11445,
  Ground-based and Airborne Telescopes VIII, ed. H.~K. {Marshall},
  J.~{Spyromilio}, \& T.~{Usuda}, 114450J, \dodoi{10.1117/12.2561419}

\bibitem[{{Herbst} {et~al.}(2022){Herbst}, {Bizenberger}, {Blanc}, {Briegel},
  {Case}, {Drory}, {Feger}, {Froning}, {Gaessler}, {H{\"a}berle}, {Hebert},
  {Kollmeier}, {Konidaris}, {Kuhlberg}, {Lanz}, {Mathar}, {Mohr}, {Pak},
  {Ram{\'\i}rez}, {Ritz}, {Rohloff}, {S{\'a}nchez-Gallego}, {Stepie{\'n}},
  {Wachter}, {Zhelem}, \& {Bizyaev}}]{Herbst+2022}
{Herbst}, T.~M., {Bizenberger}, P., {Blanc}, G., {et~al.} 2022, in Society of
  Photo-Optical Instrumentation Engineers (SPIE) Conference Series, Vol. 12182,
  Ground-based and Airborne Telescopes IX, ed. H.~K. {Marshall},
  J.~{Spyromilio}, \& T.~{Usuda}, 121821Q, \dodoi{10.1117/12.2629354}

\bibitem[{{Herbst} {et~al.}(2024){Herbst}, {Bizenberger}, {Blanc}, {Briegel},
  {Drory}, {Froning}, {Gaessler}, {H{\"a}berle}, {Konidaris}, {Kuhlberg},
  {Lanz}, {Mathar}, {Mohr}, {Ramirez}, {Ritz}, {Rohloff},
  {S{\'a}nchez-Gallego}, {Wachter}, {Ahn}, {Besser}, {Case}, {Feger}, {Hebert},
  {Kollmeier}, {Pak}, {Rix}, {Robertson}, {St{\k{e}}pie{\'n}}, \&
  {Zhelem}}]{Herbst+2024}
{Herbst}, T.~M., {Bizenberger}, P., {Blanc}, G.~A., {et~al.} 2024, \aj, 168,
  267, \dodoi{10.3847/1538-3881/ad794810.1134/S1063772908070044}

\bibitem[{{Husemann} {et~al.}(2013){Husemann}, {Jahnke}, {S{\'a}nchez},
  {Barrado}, {Bekerait{\.{e}}}, {Bomans}, {Castillo-Morales},
  {Catal{\'a}n-Torrecilla}, {Cid Fernandes}, {Falc{\'o}n-Barroso},
  {Garc{\'\i}a-Benito}, {Gonz{\'a}lez Delgado}, {Iglesias-P{\'a}ramo},
  {Johnson}, {Kupko}, {L{\'o}pez-Fernandez}, {Lyubenova}, {Marino}, {Mast},
  {Miskolczi}, {Monreal-Ibero}, {Gil de Paz}, {P{\'e}rez}, {P{\'e}rez},
  {Rosales-Ortega}, {Ruiz-Lara}, {Schilling}, {van de Ven}, {Walcher}, {Alves},
  {de Amorim}, {Backsmann}, {Barrera-Ballesteros}, {Bland-Hawthorn}, {Cortijo},
  {Dettmar}, {Demleitner}, {D{\'\i}az}, {Enke}, {Florido}, {Flores}, {Galbany},
  {Gallazzi}, {Garc{\'\i}a-Lorenzo}, {Gomes}, {Gruel}, {Haines}, {Holmes},
  {Jungwiert}, {Kalinova}, {Kehrig}, {Kennicutt}, {Klar}, {Lehnert},
  {L{\'o}pez-S{\'a}nchez}, {de Lorenzo-C{\'a}ceres}, {M{\'a}rmol-Queralt{\'o}},
  {M{\'a}rquez}, {Mendez-Abreu}, {Moll{\'a}}, {del Olmo}, {Meidt}, {Papaderos},
  {Puschnig}, {Quirrenbach}, {Roth}, {S{\'a}nchez-Bl{\'a}zquez}, {Spekkens},
  {Singh}, {Stanishev}, {Trager}, {Vilchez}, {Wild}, {Wisotzki}, {Zibetti}, \&
  {Ziegler}}]{Husemann+2013}
{Husemann}, B., {Jahnke}, K., {S{\'a}nchez}, S.~F., {et~al.} 2013, \aap, 549,
  A87, \dodoi{10.1051/0004-6361/201220582}

\bibitem[{{Ibarra-Medel} {et~al.}(2022){Ibarra-Medel}, {Avila-Reese},
  {Lacerna}, {Rodr{\'\i}guez-Puebla}, {V{\'a}zquez-Mata},
  {Hern{\'a}ndez-Toledo}, \& {S{\'a}nchez}}]{Ibarra-Medel+2022a}
{Ibarra-Medel}, H., {Avila-Reese}, V., {Lacerna}, I., {et~al.} 2022, \mnras,
  510, 5676, \dodoi{10.1093/mnras/stab3765}

\bibitem[{{Ibarra-Medel} {et~al.}(2025){Ibarra-Medel}, {Negrete}, {Lacerna},
  {Hern{\'a}ndez-Toledo}, {Cortes-Su{\'a}rez}, \&
  {S{\'a}nchez}}]{Ibarra-Medel+2025}
{Ibarra-Medel}, H., {Negrete}, C.~A., {Lacerna}, I., {et~al.} 2025, \mnras,
  536, 752, \dodoi{10.1093/mnras/stae2623}

\bibitem[{{Ibarra-Medel} {et~al.}(2019){Ibarra-Medel}, {Avila-Reese},
  {S{\'a}nchez}, {Gonz{\'a}lez-Samaniego}, \&
  {Rodr{\'\i}guez-Puebla}}]{Ibarra-Medel+2019}
{Ibarra-Medel}, H.~J., {Avila-Reese}, V., {S{\'a}nchez}, S.~F.,
  {Gonz{\'a}lez-Samaniego}, A., \& {Rodr{\'\i}guez-Puebla}, A. 2019, \mnras,
  483, 4525, \dodoi{10.1093/mnras/sty3256}

\bibitem[{{Ibarra-Medel} {et~al.}(2016){Ibarra-Medel}, {S{\'a}nchez},
  {Avila-Reese}, {Hern{\'a}ndez-Toledo}, {Gonz{\'a}lez}, {Drory}, {Bundy},
  {Bizyaev}, {Cano-D{\'\i}az}, {Malanushenko}, {Pan}, {Roman-Lopes}, \&
  {Thomas}}]{Ibarra-Medel+2016}
{Ibarra-Medel}, H.~J., {S{\'a}nchez}, S.~F., {Avila-Reese}, V., {et~al.} 2016,
  \mnras, 463, 2799, \dodoi{10.1093/mnras/stw2126}

\bibitem[{{Kollmeier} {et~al.}(2026){Kollmeier}, {Rix}, {Aerts}, {Aird}, {Vera
  Alfaro}, {Almeida}, {Anderson}, {Arseneau}, {Assef}, {Aviram}, \&
  et~al.}]{Kollmeier+2026}
{Kollmeier}, J.~A., {Rix}, H.-W., {Aerts}, C., {et~al.} 2026, \aj, 171, 52,
  \dodoi{10.3847/1538-3881/ae0576}

\bibitem[{{Konidaris} {et~al.}(2020){Konidaris}, {Drory}, {Froning}, {Hebert},
  {Bilgi}, {Blanc}, {Lanz}, {Hull}, {Kollmeier}, {Ramirez}, {Wachter},
  {Kreckel}, {Pak}, {Pellegrini}, {Almeida}, {Case}, {Zhelem}, {Feger},
  {Lawrence}, {Lesser}, {Herbst}, {Sanchez-Gallego}, {Bershady},
  {Chattopadhyay}, {Hauser}, {Smith}, {Wolf}, \& {Yan}}]{Konidaris+2020}
{Konidaris}, N.~P., {Drory}, N., {Froning}, C.~S., {et~al.} 2020, in Society of
  Photo-Optical Instrumentation Engineers (SPIE) Conference Series, Vol. 11447,
  Ground-based and Airborne Instrumentation for Astronomy VIII, ed. C.~J.
  {Evans}, J.~J. {Bryant}, \& K.~{Motohara}, 1144718,
  \dodoi{10.1117/12.2557565}

\bibitem[{{Kreckel} {et~al.}(2024){Kreckel}, {Egorov}, {Egorova}, {Blanc},
  {Drory}, {Kounkel}, {M{\'e}ndez-Delgado}, {Rom{\'a}n-Z{\'u}{\~n}iga},
  {S{\'a}nchez}, {Stringfellow}, {Stutz}, {Zari}, {Barrera-Ballesteros},
  {Bizyaev}, {Brownstein}, {Congiu}, {Fern{\'a}ndez-Trincado}, {Garc{\'\i}a},
  {Hillenbrand}, {Ibarra-Medel}, {Jin}, {Johnston}, {Jones}, {Kim},
  {Kollmeier}, {Kong}, {Krishnarao}, {Kumari}, {Li}, {Long},
  {Mata-S{\'a}nchez}, {Mej{\'\i}a-Narv{\'a}ez}, {Popa}, {Rix}, {Sattler},
  {Serna}, {Singh}, {S{\'a}nchez-Gallego}, {Wofford}, \& {Wong}}]{Kreckel+2024}
{Kreckel}, K., {Egorov}, O.~V., {Egorova}, E., {et~al.} 2024, \aap, 689, A352,
  \dodoi{10.1051/0004-6361/202449943}

\bibitem[{{Lacerda} {et~al.}(2022){Lacerda}, {S{\'a}nchez},
  {Mej{\'\i}a-Narv{\'a}ez}, {Camps-Fari{\~n}a}, {Espinosa-Ponce},
  {Barrera-Ballesteros}, {Ibarra-Medel}, \& {Lugo-Aranda}}]{Lacerda+2022}
{Lacerda}, E. A.~D., {S{\'a}nchez}, S.~F., {Mej{\'\i}a-Narv{\'a}ez}, A.,
  {et~al.} 2022, NewAstronomy, 97, 101895, \dodoi{10.1016/j.newast.2022.101895}

\bibitem[{{Lanz} {et~al.}(2022){Lanz}, {Crane}, {Herbst}, {Bizenberger},
  {Blanc}, {Drory}, {Froning}, {Gaessler}, {Hebert}, {Hull}, {Konidaris},
  {Ram{\'\i}rez}, \& {Wachter}}]{Lanz+2022}
{Lanz}, A.~E., {Crane}, J.~D., {Herbst}, T.~M., {et~al.} 2022, in Society of
  Photo-Optical Instrumentation Engineers (SPIE) Conference Series, Vol. 12184,
  Ground-based and Airborne Instrumentation for Astronomy IX, ed. C.~J.
  {Evans}, J.~J. {Bryant}, \& K.~{Motohara}, 121845T,
  \dodoi{10.1117/12.2629633}

\bibitem[{{Lasker} {et~al.}(1990){Lasker}, {Sturch}, {McLean}, {Russell},
  {Jenkner}, \& {Shara}}]{DSS}
{Lasker}, B.~M., {Sturch}, C.~R., {McLean}, B.~J., {et~al.} 1990, \aj, 99,
  2019, \dodoi{10.1086/115483}

\bibitem[{{Law} {et~al.}(2016){Law}, {Cherinka}, {Yan}, {Andrews}, {Bershady},
  {Bizyaev}, {Blanc}, {Blanton}, {Bolton}, {Brownstein}, {Bundy}, {Chen},
  {Drory}, {D'Souza}, {Fu}, {Jones}, {Kauffmann}, {MacDonald}, {Masters},
  {Newman}, {Parejko}, {S{\'a}nchez-Gallego}, {S{\'a}nchez}, {Schlegel},
  {Thomas}, {Wake}, {Weijmans}, {Westfall}, \& {Zhang}}]{Law+2016}
{Law}, D.~R., {Cherinka}, B., {Yan}, R., {et~al.} 2016, \aj, 152, 83,
  \dodoi{10.3847/0004-6256/152/4/83}

\bibitem[{{Leroy} {et~al.}(2021){Leroy}, {Schinnerer}, {Hughes}, {Rosolowsky},
  {Pety}, {Schruba}, {Usero}, {Blanc}, {Chevance}, {Emsellem}, {Faesi},
  {Herrera}, {Liu}, {Meidt}, {Querejeta}, {Saito}, {Sandstrom}, {Sun},
  {Williams}, {Anand}, {Barnes}, {Behrens}, {Belfiore}, {Benincasa},
  {Be{\v{s}}li{\'c}}, {Bigiel}, {Bolatto}, {den Brok}, {Cao}, {Chandar},
  {Chastenet}, {Chiang}, {Congiu}, {Dale}, {Deger}, {Eibensteiner}, {Egorov},
  {Garc{\'\i}a-Rodr{\'\i}guez}, {Glover}, {Grasha}, {Henshaw}, {Ho}, {Kepley},
  {Kim}, {Klessen}, {Kreckel}, {Koch}, {Kruijssen}, {Larson}, {Lee}, {Lopez},
  {Machado}, {Mayker}, {McElroy}, {Murphy}, {Ostriker}, {Pan}, {Pessa},
  {Puschnig}, {Razza}, {S{\'a}nchez-Bl{\'a}zquez}, {Santoro}, {Sardone},
  {Scheuermann}, {Sliwa}, {Sormani}, {Stuber}, {Thilker}, {Turner}, {Utomo},
  {Watkins}, \& {Whitmore}}]{Leroy2021}
{Leroy}, A.~K., {Schinnerer}, E., {Hughes}, A., {et~al.} 2021, \apjs, 257, 43,
  \dodoi{10.3847/1538-4365/ac17f3}

\bibitem[{{M{\'a}rmol-Queralt{\'o}} {et~al.}(2011){M{\'a}rmol-Queralt{\'o}},
  {S{\'a}nchez}, {Marino}, {Mast}, {Viironen}, {Gil de Paz},
  {Iglesias-P{\'a}ramo}, {Rosales-Ortega}, \& {Vilchez}}]{Marmol+2011}
{M{\'a}rmol-Queralt{\'o}}, E., {S{\'a}nchez}, S.~F., {Marino}, R.~A., {et~al.}
  2011, \aap, 534, A8, \dodoi{10.1051/0004-6361/201117032}

\bibitem[{{McLean} {et~al.}(2000){McLean}, {Greene}, {Lattanzi}, \&
  {Pirenne}}]{DSS2a}
{McLean}, B.~J., {Greene}, G.~R., {Lattanzi}, M.~G., \& {Pirenne}, B. 2000, in
  Astronomical Society of the Pacific Conference Series, Vol. 216, Astronomical
  Data Analysis Software and Systems IX, ed. N.~{Manset}, C.~{Veillet}, \&
  D.~{Crabtree}, 145

\bibitem[{{P{\'e}rez} {et~al.}(2013){P{\'e}rez}, {Cid Fernandes}, {Gonz{\'a}lez
  Delgado}, {Garc{\'{\i}}a-Benito}, {S{\'a}nchez}, {Husemann}, {Mast},
  {Rod{\'o}n}, {Kupko}, {Backsmann}, {de Amorim}, {van de Ven}, {Walcher},
  {Wisotzki}, {Cortijo-Ferrero}, \& {CALIFA Collaboration}}]{Perez+2013}
{P{\'e}rez}, E., {Cid Fernandes}, R., {Gonz{\'a}lez Delgado}, R.~M., {et~al.}
  2013, \apjl, 764, L1, \dodoi{10.1088/2041-8205/764/1/L1}

\bibitem[{{Perruchot} {et~al.}(2018){Perruchot}, {Guy}, {Le Guillou}, {Blanc},
  {Ronayette}, {R{\'e}gal}, {Castagnoli}, {Sepulveda}, {Le Van Suu}, {Jullo},
  {Cuby}, {Karkar}, {Ghislain}, {Repain}, {Carton}, {Magneville}, {Ealet},
  {Escoffier}, {Secroun}, {Cousinou}, {Honscheid}, {Elliot}, {Jelinsky},
  {Brooks}, \& {Tarl{\`e}}}]{Perruchot+2018}
{Perruchot}, S., {Guy}, J., {Le Guillou}, L., {et~al.} 2018, in Society of
  Photo-Optical Instrumentation Engineers (SPIE) Conference Series, Vol. 10702,
  Ground-based and Airborne Instrumentation for Astronomy VII, ed. C.~J.
  {Evans}, L.~{Simard}, \& H.~{Takami}, 107027K, \dodoi{10.1117/12.2311996}

\bibitem[{{Richardson}(1972)}]{Richardson+72}
{Richardson}, W.~H. 1972, Journal of the Optical Society of America
  (1917-1983), 62, 55

\bibitem[{{S{\'a}nchez}(2006)}]{Sanchez+2006}
{S{\'a}nchez}, S.~F. 2006, Astronomische Nachrichten, 327, 850,
  \dodoi{10.1002/asna.200610643}

\bibitem[{{S{\'a}nchez}(2020)}]{Sanchez+2020}
---. 2020, \araa, 58, 99, \dodoi{10.1146/annurev-astro-012120-013326}

\bibitem[{{S{\'a}nchez} {et~al.}(2023){S{\'a}nchez}, {Galbany}, {Walcher},
  {Garc{\'\i}a-Benito}, \& {Barrera-Ballesteros}}]{Sanchez+2023}
{S{\'a}nchez}, S.~F., {Galbany}, L., {Walcher}, C.~J., {Garc{\'\i}a-Benito},
  R., \& {Barrera-Ballesteros}, J.~K. 2023, \mnras, 526, 5555,
  \dodoi{10.1093/mnras/stad3119}

\bibitem[{{S{\'a}nchez} {et~al.}(2012){S{\'a}nchez}, {Kennicutt}, {Gil de Paz},
  {van de Ven}, {V{\'\i}lchez}, {Wisotzki}, {Walcher}, {Mast}, {Aguerri},
  {Albiol-P{\'e}rez}, {Alonso-Herrero}, {Alves}, {Bakos}, {Bart{\'a}kov{\'a}},
  {Bland-Hawthorn}, {Boselli}, {Bomans}, {Castillo-Morales}, {Cortijo-Ferrero},
  {de Lorenzo-C{\'a}ceres}, {Del Olmo}, {Dettmar}, {D{\'\i}az}, {Ellis},
  {Falc{\'o}n-Barroso}, {Flores}, {Gallazzi}, {Garc{\'\i}a-Lorenzo},
  {Gonz{\'a}lez Delgado}, {Gruel}, {Haines}, {Hao}, {Husemann},
  {Igl{\'e}sias-P{\'a}ramo}, {Jahnke}, {Johnson}, {Jungwiert}, {Kalinova},
  {Kehrig}, {Kupko}, {L{\'o}pez-S{\'a}nchez}, {Lyubenova}, {Marino},
  {M{\'a}rmol-Queralt{\'o}}, {M{\'a}rquez}, {Masegosa}, {Meidt},
  {Mendez-Abreu}, {Monreal-Ibero}, {Montijo}, {Mour{\~a}o}, {Palacios-Navarro},
  {Papaderos}, {Pasquali}, {Peletier}, {P{\'e}rez}, {P{\'e}rez}, {Quirrenbach},
  {Rela{\~n}o}, {Rosales-Ortega}, {Roth}, {Ruiz-Lara},
  {S{\'a}nchez-Bl{\'a}zquez}, {Sengupta}, {Singh}, {Stanishev}, {Trager},
  {Vazdekis}, {Viironen}, {Wild}, {Zibetti}, \& {Ziegler}}]{Sanchez+2012}
{S{\'a}nchez}, S.~F., {Kennicutt}, R.~C., {Gil de Paz}, A., {et~al.} 2012,
  \aap, 538, A8, \dodoi{10.1051/0004-6361/201117353}

\bibitem[{{S{\'a}nchez} {et~al.}(2016{\natexlab{a}}){S{\'a}nchez}, {P{\'e}rez},
  {S{\'a}nchez-Bl{\'a}zquez}, {Gonz{\'a}lez}, {Ros{\'a}les-Ortega},
  {Cano-D{\'\i}az}, {L{\'o}pez-Cob{\'a}}, {Marino}, {Gil de Paz}, {Moll{\'a}},
  {L{\'o}pez-S{\'a}nchez}, {Ascasibar}, \&
  {Barrera-Ballesteros}}]{Sanchez+2016a}
{S{\'a}nchez}, S.~F., {P{\'e}rez}, E., {S{\'a}nchez-Bl{\'a}zquez}, P., {et~al.}
  2016{\natexlab{a}}, RevMex, 52, 21, \dodoi{10.48550/arXiv.1509.08552}

\bibitem[{{S{\'a}nchez} {et~al.}(2016{\natexlab{b}}){S{\'a}nchez}, {P{\'e}rez},
  {S{\'a}nchez-Bl{\'a}zquez}, {Garc{\'\i}a-Benito}, {Ibarra-Mede},
  {Gonz{\'a}lez}, {Rosales-Ortega}, {S{\'a}nchez-Menguiano}, {Ascasibar},
  {Bitsakis}, {Law}, {Cano-D{\'\i}az}, {L{\'o}pez-Cob{\'a}}, {Marino}, {Gil de
  Paz}, {L{\'o}pez-S{\'a}nchez}, {Barrera-Ballesteros}, {Galbany}, {Mast},
  {Abril-Melgarejo}, \& {Roman-Lopes}}]{Sanchez+2016b}
---. 2016{\natexlab{b}},RevMex, 52, 171, \dodoi{10.48550/arXiv.1602.01830}

\bibitem[{{S{\'a}nchez} {et~al.}(2022){S{\'a}nchez}, {Barrera-Ballesteros},
  {Lacerda}, {Mej{\'\i}a-Narvaez}, {Camps-Fari{\~n}a}, {Bruzual},
  {Espinosa-Ponce}, {Rodr{\'\i}guez-Puebla}, {Calette}, {Ibarra-Medel},
  {Avila-Reese}, {Hernandez-Toledo}, {Bershady}, {Cano-Diaz}, \&
  {Munguia-Cordova}}]{Sanchez+2022}
{S{\'a}nchez}, S.~F., {Barrera-Ballesteros}, J.~K., {Lacerda}, E., {et~al.}
  2022, \apjs, 262, 36, \dodoi{10.3847/1538-4365/ac7b8f}

\bibitem[{{S{\'a}nchez} {et~al.}(2025){S{\'a}nchez}, {Mej{\'\i}a-Narv{\'a}ez},
  {Egorov}, {Kreckel}, {Drory}, {Blanc}, {M{\'e}ndez-Delgado},
  {Barrera-Ballesteros}, {Ibarra-Medel}, {Bizyaev}, {Garc{\'\i}a}, {Wofford},
  \& {Lugo-Aranda}}]{Sanchez+2024}
{S{\'a}nchez}, S.~F., {Mej{\'\i}a-Narv{\'a}ez}, A., {Egorov}, O.~V., {et~al.}
  2025, \aj, 169, 52, \dodoi{10.3847/1538-3881/ad93bb}

\bibitem[{{STScI}(2020)}]{DSS2b}
{STScI}. 2020, Digitized Sky Survey,  IPAC, \dodoi{10.26131/IRSA441}

\bibitem[{{Tritton}(1978)}]{Tritton+1978}
{Tritton}, S.~B. 1978, PASA, 3, 206, \dodoi{10.1017/S1323358000024565}

\bibitem[{{van der Marel}(1995)}]{vanderMarel1995}
{van der Marel}, R.~P. 1995, in Calibrating Hubble Space Telescope. Post
  Servicing Mission, ed. A.~P. {Koratkar} \& C.~{Leitherer}, 94,
  \dodoi{10.48550/arXiv.astro-ph/9505095}

\bibitem[{{van der Marel} {et~al.}(1997){van der Marel}, {de Zeeuw}, \&
  {Rix}}]{vanderMarel1997}
{van der Marel}, R.~P., {de Zeeuw}, P.~T., \& {Rix}, H.-W. 1997, \apj, 488,
  119, \dodoi{10.1086/304690}

\bibitem[{van~der Walt {et~al.}(2014)van~der Walt, Sch\"{o}nberger,
  Nunez-Iglesias, Boulogne, Warner, Yager, Gouillart, \& Yu}]{vanderWalt2014}
van~der Walt, S., Sch\"{o}nberger, J.~L., Nunez-Iglesias, J., {et~al.} 2014,
  PeerJ, 2, e453, \dodoi{10.7717/peerj.453}

\end{thebibliography}

%\appendix

% Don't change these lines
%\bsp	% typesetting comment
\label{lastpage}
\end{document}